\documentclass[reprint,aps,amsmath,amssymb,showpacs,preprintnumbers,superscriptaddress]{revtex4-1}
\usepackage{graphicx}

\begin{document}

\preprint{TTK-11-30, WUB/11-08}

\title{Hadronic top-quark pair production in association with two jets \\
at Next-to-Leading Order QCD}

\author{G. Bevilacqua}
\affiliation{Institut f\"ur Theoretische Teilchenphysik und
Kosmologie, RWTH Aachen University, D-52056 Aachen, Germany}
\author{M. Czakon}
\affiliation{Institut f\"ur Theoretische Teilchenphysik und
  Kosmologie, RWTH Aachen University, D-52056 Aachen, Germany}
\author{C. G. Papadopoulos}
\affiliation{Institute of Nuclear Physics, NCSR Demokritos, GR-15310
             Athens, Greece}
\author{M. Worek}
\affiliation{Fachbereich C, Bergische Universit\"at Wuppertal, D-42097
             Wuppertal, Germany}

\date{\today}

\begin{abstract}
We report on the calculation of the next-to-leading order QCD
corrections to  the production of $t\bar{t}$ pairs in association
with two hard jets at the  Fermilab TeVatron and CERN Large Hadron
Collider.  Results for the integrated and differential cross
sections are given.  The corrections with respect to leading order
are negative and moderate.  A study of the scale dependence of our NLO
predictions indicates that the residual theoretical uncertainty, due to
higher order corrections, is $21\%$ for the TeVatron and $15\%$ for the
LHC.  In case of the TeVatron, the
forward-backward asymmetry of the top quark is calculated for the
first time at next-to-leading order. With the inclusive selection of
cuts,  this
asymmetry  amounts  to ${\cal A}_{\rm FB, LO}^{\rm t} = -10.3\% $ at leading
order and  ${\cal A}_{\rm FB, NLO}^{\rm t}  = -4.6 \%$ at next-to-leading
order. All results presented in this paper have been obtained with the help 
of the \textsc{Helac-Nlo} package.

\end{abstract}


\maketitle

%
\section{Introduction}
%
Top quark production in association with two jets constitutes an
important background for many new particle  searches at the Tevatron
and at the LHC. Prominent examples include   searches for  Higgs boson
decays $H \to WW^{*}$  and $H\to b\bar{b}$,  where the Higgs boson is
produced via weak boson fusion or via  an associated  production
with  a $t\bar{t}$ pair respectively.  Higgs boson production in association
with top quarks has a strongly decreasing  cross section
with increasing Higgs boson mass,  which makes the process useful only 
in  the low mass range, $m_H \le 135 ~{\rm GeV}$, when Higgs boson
decays in $b\bar{b}$ are important.  This potential discovery
channel
gives a unique access  to the top and bottom Yukawa couplings. 
Combined analyzes of ATLAS
\cite{Aad:2009wy} and CMS \cite{Ball:2007zza} have shown that a
$3\sigma$ evidence of  the signal above the dominant
$t\bar{t}b\bar{b}$ and $t\bar{t}jj$ backgrounds can be obtained for
$M_H \le 130$ GeV at the LHC, if enough luminosity, ${\cal L} = {\rm
  60 ~fb^{-1}}$,  is collected. However, a reconstruction of the $H\to
b\bar{b}$ mass peak is difficult because of an identification
problem in the signal.  The $b\bar{b}$ pair can be
chosen incorrectly, due to  a lack of a distinctive kinematic feature of
jets from the Higgs boson decay. Moreover, the b-tagging efficiency takes 
a crucial part in this kind of analyzes, since two b-jets for a Higgs 
boson candidate can arise from mistagged light jets.  Therefore, a very 
precise knowledge of the backgrounds is necessary, if the $t\bar{t}H$ 
channel is to be of any use.  

On the other hand, a detailed ATLAS examination of the vector  boson
fusion channel at the LHC,  $pp\to Hjj\to W^+W^- jj\to \ell \nu\ell
\nu jj$,  has shown  that a significance larger than $3\sigma$ can be
obtained for a luminosity of ${\cal L} = 10 ~{\rm fb^{-1}}$ in the
Higgs boson mass range above  $m_H \sim$  130 GeV, {\it i.e.} when ${\cal
  BR}(H \to WW^{*})$ is large enough \cite{Asai:2004ws}.  When
combining this channel with the $\ell\nu jj$ mode, one can even obtain
a $\sim 5\sigma$ significance for the above mass range. In fact, the
$pp\to Hjj \to WW jj$ channel becomes very powerful at higher Higgs
boson masses. One can arrive at a signal  significance that is above
$5\sigma$ in a wide mass range, namely from $140 ~{\rm GeV}$ up to $190
~{\rm GeV}$, for the same luminosity. Although feasible and
competitive,  this channel might prove to be rather difficult since
the Higgs boson mass peak cannot be directly
reconstructed. Consequently,  one cannot measure  background processes 
from the side bands. The most  important background processes, $t\bar{t}jj$
and QCD $W^+W^-jj$, need to  be known, therefore,  as precisely as
possible. 

Meanwhile at the TeVatron,  with up to $7.1 ~{\rm fb^{-1}}$ of data
analyzed at CDF, and up to $8.2 ~{\rm fb^{-1}}$ at D0, the  Higgs
boson mass range of  $158 ~{\rm GeV} < m_H < 173 ~{\rm GeV}$ has been
excluded  \cite{Aaltonen:2011gs}.   Very recently, the LHC experiments
have presented their Higgs boson search results based on $1 ~{\rm fb^{-1}}$
of analyzed data.  CMS has  excluded the Standard Model Higgs boson in the
$149 ~{\rm GeV} - 206$ GeV and $300 ~{\rm GeV} - 440$ GeV windows
\cite{CMS}, while ATLAS has excluded the $155 ~{\rm GeV} - 190$ GeV
and $295 ~{\rm GeV} - 450$ GeV windows \cite{ATLAS}.   The low mass
exclusion is dominated by the search of the  $H \to W^+W^-\to \ell
\nu\ell\nu$ final state, while the high mass one  is dominated by
$H \to ZZ$ with three different combinations of Z decays.  Considering
these recent experimental results, an  excellent understanding of
lighter Higgs boson  search channels is more timely than
ever. 

Apart from its significance as a background process to various new
physics searches it turns out that  $t\bar{t}jj$ production can also
be an important signal process.  With the total integrated luminosity
accumulated so far at the TeVatron around $3000$ events should have
been collected so far by each experiment.  As for the LHC, in June of
this year  both experiments ATLAS and CMS reached $1 ~{\rm fb^{-1}}$ of
collected  data. This can be directly translated to  about $10 000$
 events already stored. Since they are in the  process of collecting
data, the $t\bar{t}jj$ process can soon   become of great importance
for a more precise  understanding of the top quark
events topology. 

However, it is a well known fact that for processes involving strongly
interacting  particles, the leading order (LO) cross sections are
affected by large  uncertainties arising from higher order
corrections. If, at least, next-to-leading order (NLO) QCD
corrections to these processes are included, total cross sections
can be defined in a reliable way. The NLO QCD corrections to $pp\to
t\bar{t}b\bar{b}$ production have already been calculated by two
independent groups
\cite{Bredenstein:2009aj,Bevilacqua:2009zn,Bredenstein:2010rs}, and
perfect agreement has been found.  In addition, NLO corrections to the
remaining two background processes, namely  $pp\to W^+W^-jj$
\cite{Melia:2011dw} and $pp\to t\bar{t}jj$ \cite{Bevilacqua:2010ve}
have recently been evaluated.  In this paper, we have extend our
previous study on the NLO QCD corrections  to the  $pp\to t\bar{t}jj$
production at the LHC, where the  integrated cross section together
with the scale dependence of the total  cross section for one particular
set of  cuts for the center of mass system energy of $\sqrt{s}=14
~{\rm TeV}$ have been  studied.  Moreover, we have presented there
three differential distributions, the invariant mass distribution of
the first and the second hardest jet together with the transverse
momentum  distributions of the first and the second hard jet.  We
supplement this discussion here with more results. In particular,
results for the TeVatron case are going to be  presented for the first
time. In case of the LHC,  we use the current center of mass  system energy of
$\sqrt{s}= 7 ~{\rm GeV}$.  Additionally to the integrated cross
sections for different setups, various  differential distributions are
given. 

The article is organized as follows. In Section \ref{sec:1} we briefly
summarize the methods we use to obtain our NLO QCD results. In Section
\ref{sec:2} numerical results are presented, first for the TeVatron
and later on for the  LHC case.  We give our conclusions in Section
\ref{sec:3}. Finally, in an Appendix we give values of squared
matrix elements for the virtual corrections for one phase space point. 
 
%
\section{Framework of the calculation}
\label{sec:1}
The calculation of NLO corrections to  hadronic top pair production in
association with two jets proceeds along the same lines as our earlier
work on $pp\rightarrow t\bar{t}b\bar{b} $ \cite{Bevilacqua:2009zn},
$pp\to t\bar{t}jj$ \cite{Bevilacqua:2010ve} and
$pp(p\bar{p})\rightarrow  t\bar{t}\rightarrow
W^+W^-b\bar{b}\rightarrow e^+\nu_{e}\mu^-\bar{\nu}_{\mu}b\bar{b}$
\cite{Bevilacqua:2010qb}. The methods developed there have  therefore
been straightforwardly adapted and only need a brief recollection
here.  Let us emphasize, that all  parts of the calculation are
performed fully numerically in a completely automatic manner.
%
\subsection{Tree level contributions}
\begin{figure}
\includegraphics[width=0.48\textwidth]{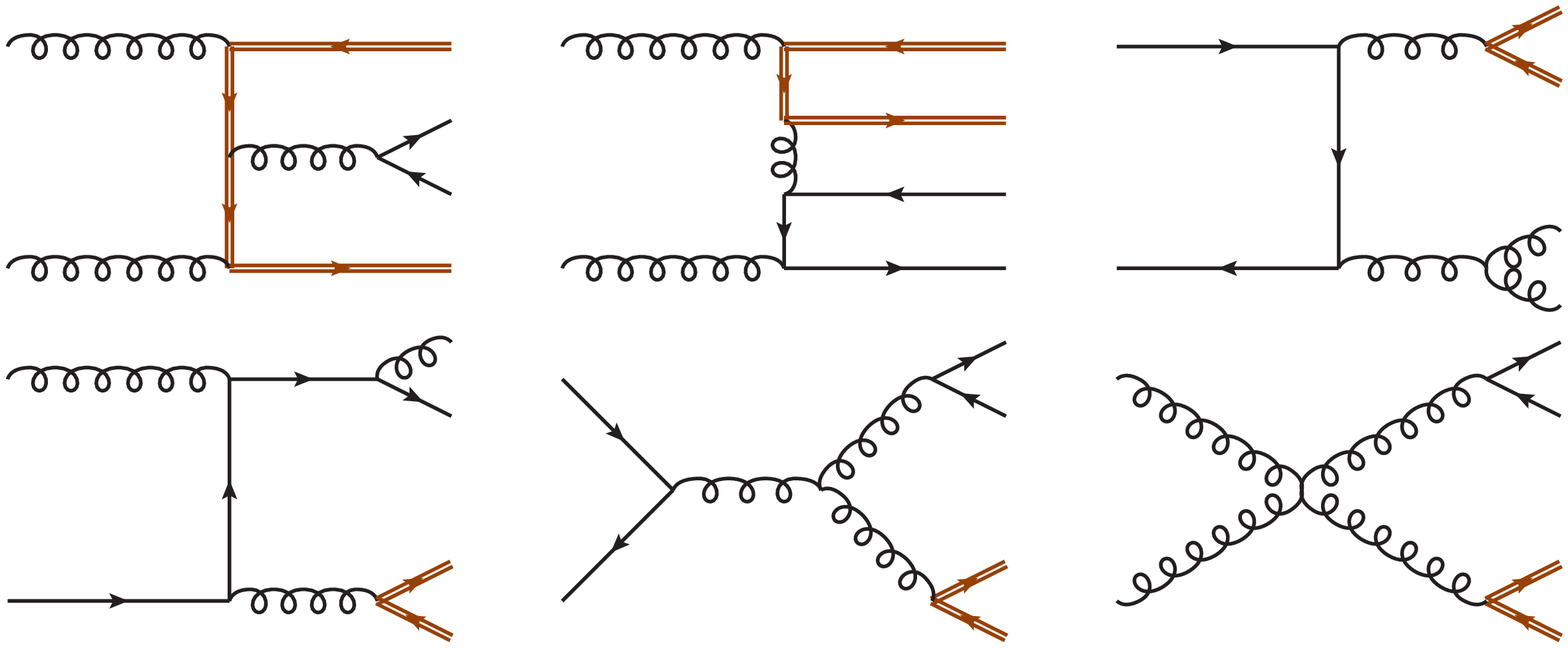}
\caption{\it \label{fig:diagrams-lo} A representative set of Feynman
  diagrams contributing to the LO hadronic $t\bar{t}jj$ production at
  ${\cal{O}}(\alpha_{s}^4)$. Double brown lines correspond to top
  quarks, single lines to light quarks and wiggly ones to gluons.}
\end{figure}

%
At tree level the $t\bar{t}$ production in association with two jets
proceeds via the scattering of two gluons, two quarks or quark and
gluon. All contributions of the  order of ${\cal{O}}(\alpha_{s}^4)$
can be grouped into eight easily identifiable topologies presented in the
Table \ref{tab:lo}.
%
\begin{table}
  \caption{\it \label{tab:lo} Partonic subprocesses contributing to
    the leading order process  $pp(p\bar{p})\to t\bar{t}jj$  at
    ${\cal{O}}(\alpha_{s}^4)$. The number of Feynman diagrams
    corresponding to these subprocesses is also shown. }
\begin{ruledtabular}
  \begin{tabular}{cc}
   \textsc{Partonic }    & \textsc{Number Of Feynman}\\
  \textsc{Subprocess}    & \textsc{Diagrams}\\
 &\\
\hline
$gg \to  t\bar{t} gg$  & 123 \\ 
$gg  \to t\bar{t} q\bar{q}$ &  36 \\
$q\bar{q} \to  t\bar{t}gg $& 36 \\
$gq \to   t\bar{t} qg $&  36 \\
$qg  \to  t\bar{t} qg $&  36\\
$qq^{\prime} \to t\bar{t} qq^{\prime}$  & 7\\
$q\bar{q}  \to t\bar{t}q^{\prime}\bar{q}^{\prime} $& 7\\
$q\bar{q} \to  t\bar{t}q\bar{q} $&  14 \\
  \end{tabular}
\end{ruledtabular}
\end{table}
%
Both $q$ and ${q}^{\prime}$ span all quarks and anti-quarks.  Also
shown  are the numbers of Feynman diagrams  corresponding to these
subprocesses. We emphasis here that we do not use Feynman diagrams
in our calculations. Instead, matrix elements and total  cross
sections are obtained with an iterative algorithm based on the
Dyson-Schwinger equations \cite{Draggiotis:1998gr,Draggiotis:2002hm}.
Nevertheless, we present the number of Feynman diagrams as a  measure of
complexity. All partonic channels  have been calculated  with the
\textsc{Helac-Dipoles} \cite{Czakon:2009ss} package.  An independent
calculation using the \textsc{Helac-Phegas} \cite{Kanaki:2000ey,
  Papadopoulos:2000tt,Cafarella:2007pc} event generator has been
performed.  Let us stress that \textsc{Helac-Phegas} has been
already extensively used and tested, see {\it e.g.}
\cite{Gleisberg:2003bi,Papadopoulos:2005ky,Alwall:2007fs,
  Englert:2008tn,Actis:2010gg,Calame:2011zq}.     Moreover, both
libraries compute relevant contributions in a very efficient way,
using off-shell recursive equations and avoiding a multiple evaluation of
recurring building blocks.  In every case  we have found full
numerical agreement.  The optimization and phase space integration was
executed with the help of \textsc{Parni} \cite{vanHameren:2007pt} and
\textsc{Kaleu} \cite{vanHameren:2010gg}.  A few examples of LO
graphs are shown in  Figure \ref{fig:diagrams-lo}. 
%
%
\subsection{Virtual corrections}
\begin{figure}
\includegraphics[width=0.48\textwidth]{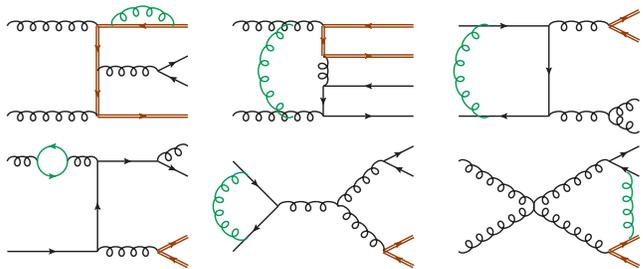}
\caption{\it \label{fig:diagrams-nlo} A representative set of Feynman
  diagrams contributing to the virtual corrections to hadronic
  $t\bar{t}jj$ production at ${\cal{O}}(\alpha_{s}^5)$. Double brown
  lines correspond to top quarks, single lines to light quarks and
  wiggly ones to gluons. }
\end{figure}

%
The virtual corrections are obtained from the interference  of the sum
of all one-loop diagrams  with the Born amplitude.    One can classify
them into  self-energy, vertex, box-type, pentagon-type and
hexagon-type corrections. In Table \ref{tab:virtual}  the  number of
one-loop Feynman diagrams for the $pp(p\bar{p})\to
t\bar{t}jj$ process at ${\cal{O}}(\alpha_{s}^5)$ as obtained with
\textsc{Qgraph} \cite{Nogueira:1991ex} is  given.   Again they 
are shown as a  measure of complexity of our calculation since we
are free from the task  of computing Feynman  graphs for a given process.
Typical examples of
virtual graphs are shown in Figure \ref{fig:diagrams-nlo}. 
The virtual corrections
are calculated in $d=4-2\epsilon$ dimensions in  the 't Hooft-Veltman
\cite{'tHooft:1972fi} version of the dimensional regularization
scheme.  The singularities stemming from infrared divergent pieces are
canceled by the respective poles in the integrated counter terms of
the dipole subtraction  approach. The finite contributions of the loop
diagrams are evaluated numerically in  $d=4$ dimension. We monitor the
numerical stability by checking Ward identities at every phase space
point.  The events which violate gauge invariance are not
discarded from the calculation of the finite part, but rather
recalculated with higher precision.    In order to compute the
relevant contributions the \textsc{ Helac-1Loop} \cite{vanHameren:2009dr}
approach is used.  It is based on the \textsc{Helac-Phegas} program to
calculate all tree-order like ingredients and the OPP
\cite{Ossola:2006us} reduction method. The cut-constructible part of
the virtual amplitudes and  of the rational terms, $R_1$, are computed
using the \textsc{CutTools} \cite{Ossola:2007ax} code. The remaining
part of the rational term, $R_2$, is obtained by the use of extra
Feynman rules as described in \cite{Ossola:2007ax,Draggiotis:2009yb,
Garzelli:2009is,Garzelli:2010qm,Garzelli:2010fq}.
The evaluation of scalar integrals is performed with the help of
\textsc{OneLOop} \cite{vanHameren:2009dr,vanHameren:2010cp}.
Renormalization is done as usual, by evaluating tree level
diagrams with counterterms.  For our process, we chose to renormalize
the coupling in the $\overline{\rm MS}$ scheme with five active
flavors and the top quark decoupled, just as in
\cite{Bevilacqua:2009zn, Bevilacqua:2010ve}, and the mass in the
on-shell scheme (wave function renormalization is done in the on-shell
scheme as it must be).
%
\begin{table}
  \caption{\it \label{tab:virtual}  The number of one-loop Feynman diagrams 
   for the  $pp(p\bar{p})\to t\bar{t}jj$  process at 
  ${\cal{O}}(\alpha_{s}^5)$.}
\begin{ruledtabular}
  \begin{tabular}{cc}
   \textsc{Partonic }    & \textsc{Number Of Feynman}\\
  \textsc{Subprocess}    & \textsc{Diagrams}\\
 &\\
\hline
$gg \to  t\bar{t} gg$  &  4510 \\ 
$gg  \to t\bar{t} q\bar{q}$ & 1100 \\
$q\bar{q} \to  t\bar{t}gg $& 1100\\
$gq \to   t\bar{t} qg $&  1100\\
$qg  \to  t\bar{t} qg $&  1100\\
$qq^{\prime} \to t\bar{t} qq^{\prime}$  & 205 \\
$q\bar{q}  \to t\bar{t}q^{\prime}\bar{q}^{\prime} $& 205 \\
$q\bar{q} \to  t\bar{t}q\bar{q} $&  410\\
  \end{tabular}
\end{ruledtabular}
\end{table}
%
\subsection{Real corrections}
%
The real emission corrections to the LO process arise from tree level
amplitudes with one additional parton, an additional gluon, or a quark
anti-quark pair replacing a gluon.  All possible contributions of the
order  of ${\cal{O}}(\alpha_{s}^5)$ can be divided into ten
subprocesses which are given in Table \ref{tab:nlo}, where again $q$
and $q^{\prime}$ stand for up- or down-type quarks. Moreover, the number
of Feynman diagrams and the number of dipoles corresponding to these
partonic reactions are shown.
%
\begin{table}
  \caption{\it \label{tab:nlo} Partonic subprocesses contributing to
    the subtracted real emissions  at ${\cal{O}}(\alpha_{s}^5)$  for
    the $pp(p\bar{p})\to t\bar{t}jj$ process. Also, the number of Feynman
    diagrams and  the number of dipoles corresponding to these
    subprocesses are presented. }
\begin{ruledtabular}
  \begin{tabular}{ccc}
   \textsc{Partonic } & \textsc{Number Of Feynman}  & \textsc{Number Of}\\
   \textsc{Subprocess} &  \textsc{Diagrams}  & \textsc{Dipoles}\\
 &&\\
\hline
$gg \to  t\bar{t} ggg$ & 1240 & 75 \\ 
$gg  \to t\bar{t} q\bar{q}g$ & 341 & 55 \\
$q\bar{q} \to  t\bar{t}ggg $ & 341 & 75\\
$gq \to   t\bar{t} q^{\prime}\bar{q}^{\prime}q $ & 64 & 25\\
$gq \to   t\bar{t} qgg $ & 341  & 65 \\
$qg  \to  t\bar{t} q^{\prime}\bar{q}^{\prime}q $ & 64 & 25 \\
$qg  \to  t\bar{t} qgg $ & 341 & 65 \\
$qq^{\prime} \to t\bar{t} qq^{\prime}g$  & 64  & 40\\
$q\bar{q}  \to t\bar{t}q^{\prime}\bar{q}^{\prime}g $ & 64 & 35\\
$q\bar{q} \to  t\bar{t}q\bar{q}g $ & 128 & 45\\
  \end{tabular}
\end{ruledtabular}
\end{table}
We employ the dipole subtraction formalism in the form as proposed by
Catani and Seymour \cite{Catani:1996vz}, to extract the soft and
collinear infrared singularities and to combine them with the virtual
corrections.  Specifically, the formulation presented in
\cite{Catani:2002hc} for massive quarks has been used with the
extension to arbitrary helicity   eigenstates of the external partons
\cite{Czakon:2009ss},  as implemented in \textsc{Helac-Dipoles}.
After combining virtual and real corrections, singularities connected
to collinear configurations in the final state as well as soft
divergences in the initial and final states cancel for collinear-safe
observables automatically after applying a jet
algorithm. Singularities connected to collinear initial-state
splittings are removed via factorization by parton distribution
function   redefinitions. In our case the cancellation of divergences
has been  checked numerically for a few phase space points.  Let us
mention here that similarly to the LO case, the phase space integration is
executed with the help of  \textsc{Kaleu} \cite{vanHameren:2010gg}.
However, in case of subtracted real emission  a version with
dipoles channels have been used instead,  see \cite{Bevilacqua:2010qb}
for details.
%
\section {Numerical results}
\label{sec:2}
In the following we present predictions for the process  $p\bar{p}(pp)
\rightarrow t\bar{t}jj +X$ both at the TeVatron run II with
$\sqrt{s}=1.96$ TeV and at the Large Hadron Collider (LHC) with
$\sqrt{s}=7$ TeV. We have consistently employed the MSTW2008
\cite{Martin:2009iq}   set of parton distribution functions (PDFs).
In particular, we take   MSTW2008LO PDFs with 1-loop running
$\alpha_{s}$ in LO and MSTW2008NLO  PDFs with 2-loop running
$\alpha_s$ in NLO, including five active flavors.   Contributions induced
by bottom-quark density are taken into account. We use the Tevatron
average mass of the top quark $m_t = 173.3$ GeV \cite{:1900yx} as
measured by the CDF and D0 experiments. The masses of all other
quarks, including b quarks, are neglected. If not stated otherwise,
the renormalization and factorization scales are set to
$\mu_R=\mu_F=m_{t}$.   Jets are defined  via an infrared safe
algorithm, the $k_T$ \cite{Catani:1992zp,Catani:1993hr,Ellis:1993tq},
{\it anti}$-k_T$ \cite{Cacciari:2008gp} (our default)  and the
inclusive Cambridge/Aachen (C/A) algorithm \cite{Dokshitzer:1997in}.
In all cases  no clustering can take place unless two partons of
pseudorapidity
\begin{equation} 
|\eta| = -\ln \left[\tan \left(\theta/2\right)\right] < 5 \, ,
\end{equation}
where $\theta$ is the angle between the parton momentum  and the beam
axis,  are separated by less  than $R$ in the $(\eta,\phi)$
plane. Reconstructed jets are ordered in $p_T$.  Their momenta are
formed  as the four-vector sum of massless partons.  The outgoing
(anti-)top quarks  are   left on-shell with unrestricted
kinematics. They are assumed to be always  tagged.
%
\subsection{TeVatron}
%
We calculate the partonic cross sections for events with at least two hard 
jets. Jets are obtained via the {\it anti}$-k_T$ jet algorithm 
with a resolution parameter $R=0.4$ and  are required to have
\begin{equation}
p_{T_{j}} = \sqrt{p_{x_j}^2+p_{y_j}^2} > 20 ~{\rm GeV}\, , 
\end{equation}
\begin{equation}
|y_j| = \frac{1}{2}\ln\left(\frac{E_j+p_{z_j}}{E_j-p_{z_j}}\right)  < 2\, , 
\end{equation}
\begin{equation}
\Delta R_{jj} =\sqrt{\Delta\phi^2_{jj} +\Delta y^2_{jj}} > 0.4\, .
\end{equation}
We will refer to this set of cuts as {\it TeVatron default selection}.
All results presented in this Section  are given in picobarns, except
for the Table \ref{tab:tev4}, where results are shown in  femtobarns.
%
\subsubsection{Integrated cross sections}
%
\begin{table}
  \caption{\it \label{tab:tev1} Integrated cross section at LO and NLO for 
    $p\bar{p}\rightarrow t\bar{t}jj ~+ X$  production  at the TeVatron
    run II with $\sqrt{s}= 1.96  ~\textnormal{TeV}$.  Results for three
    different jet algorithms are presented,   the anti-$k_T$, $k_T$ and the
    Cambridge/Aachen jet algorithms with $R=0.4$. The scale choice is
    $\mu_R=\mu_F=m_{t}$.}
\begin{ruledtabular}
  \begin{tabular}{lcccr}
   \textsc{Cuts}    
     & $\sigma_{\rm LO}$ [pb]       &
     $\sigma_{\rm NLO}^{\rm anti-k_T}$  [pb] &   
     $\sigma_{\rm NLO}^{\rm k_T}$   [pb] &   
     $\sigma_{\rm NLO}^{\rm C/A}$   [pb] \\
 &&&&\\
\hline
    $p_{T_{j}} > 20$ GeV   &&&& \\
   $\Delta R_{jj}> 0.4$     & 0.3584(1) & 0.2709(5) & 0.2734(4) & 0.2734(4) \\
     $|y_i| < 2.0$   &&&&\\
  \end{tabular}
\end{ruledtabular}
\end{table}
\begin{table}
  \caption{\it \label{tab:tev2} Integrated cross section at LO and NLO for
    $p\bar{p}\rightarrow t\bar{t}jj ~+ X$  production  at the TeVatron
    run II with $\sqrt{s}= 1.96  ~\textnormal{TeV}$. Results   for three
    different jet algorithms are presented,  the anti-$k_T$, $k_T$ and the
    Cambridge/Aachen jet algorithms with $R=0.8$. The scale choice is
    $\mu_R=\mu_F=m_{t}$.}
\begin{ruledtabular}
  \begin{tabular}{lcccr}
  \textsc{Cuts}    & 
  $\sigma_{\rm LO}$ [pb]                &
  $\sigma_{\rm NLO}^{\rm anti-k_T}$  [pb] &   
  $\sigma_{\rm NLO}^{\rm k_T}$       [pb] &   
  $\sigma_{\rm NLO}^{\rm C/A}$       [pb] \\
 &&&&\\
\hline
   $p_{T_{j}} > 20$ GeV  &&&& \\
   $\Delta R_{jj}> 0.8$ & 
  0.2876(1)& 0.2467(3) & 0.2494(3) & 0.2491(3)  \\
  $|y_j| < 2.0$   &      &     &  &  \\
  \end{tabular}
\end{ruledtabular}
\end{table}
\begin{table}
  \caption{\it \label{tab:tev3} Integrated cross section at  NLO for 
    $p \bar{p} \rightarrow t\bar{t} jj ~+ X$  production  at the TeVatron
    run II with $\sqrt{s}= 1.96$ TeV.  Results for two
    different values of the $\alpha_{\rm max}$  parameter, jet resolution $R$ 
    and $\Delta R_{jj}$ cut for the anti-$k_T$ jet algorithm
    are presented.     The scale choice is $\mu_R=\mu_F=m_{t}$.}
\begin{ruledtabular}
  \begin{tabular}{lcr}
   \textsc{Cuts}    &  
   $\sigma_{\rm NLO}^{\rm \alpha_{max}=0.01}$ [pb]& 
   $\sigma_{\rm NLO}^{\rm \alpha_{max}=1.00}$   [pb]    \\
 &&\\
\hline
   $\Delta R_{jj}>0.4$, $R=0.4$  &  0.2709(5) &   0.2710(2) \\
   $\Delta R_{jj}>0.8$, $R=0.8$  &  0.2467(3) &   0.2466(2) \\
  \end{tabular}
\end{ruledtabular}
\end{table}
\begin{table}
  \caption{\it \label{tab:tev4}  Integrated cross section at LO and NLO for
    $p\bar{p}\rightarrow t\bar{t}jj ~+ X$  production   at the TeVatron
    run II with $\sqrt{s}= 1.96  ~\textnormal{TeV}$.  Results for the
    anti-$k_T$ jet algorithm with $R=0.4$ are presented.   In the last two
    columns the $\cal K$ factor, defined as  the ratio of the NLO cross
    section to the respective LO result, and NLO corrections in \% are given.
    The scale choice is  $\mu_R=\mu_F=m_{t}$.}
\begin{ruledtabular}
  \begin{tabular}{lcccr}
      \textsc{$p_{T_j}$ Cut}     & $\sigma_{\rm LO}$ [fb]       &
     $\sigma_{\rm NLO}^{\rm anti-k_T}$  [fb] &    ${\cal K}$ & [\%]\\
&&&&\\
\hline
     $p_{T_{j}} > 20$ GeV  & 358.4(1)  & 271.2(4)   & 0.76  & -24\\
     $p_{T_{j}} > 40$ GeV  & 79.29(3)  &  54.11(9)  & 0.68 & -32\\
     $p_{T_{j}} > 60$ GeV  & 23.85(1)  &  14.55(3)  & 0.61  & -39\\
     $p_{T_{j}} > 80$ GeV  & 8.274(3)  &  4.520(13) & 0.55  & -45 \\
  \end{tabular}
\end{ruledtabular}
\end{table}
%
\begin{figure}
\includegraphics[width=0.48\textwidth]{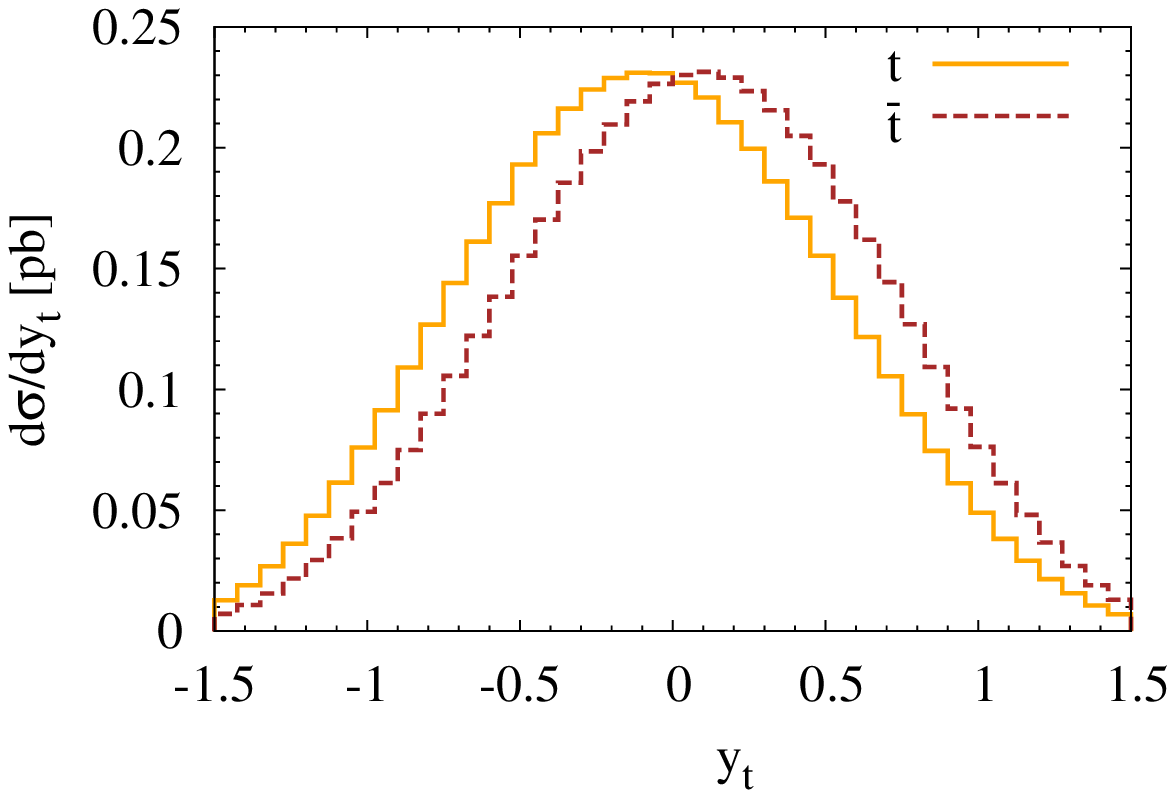}
\includegraphics[width=0.48\textwidth]{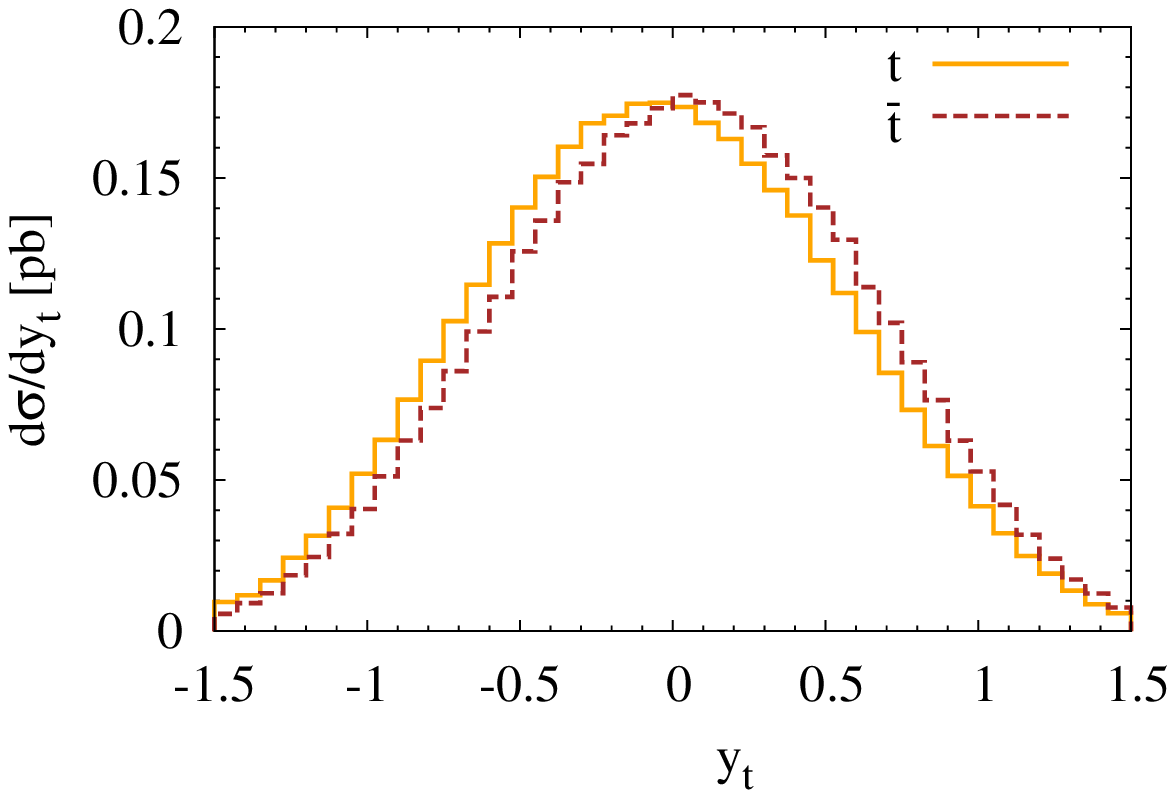}
\caption{\it \label{tev:asymmetry1} 
  Differential cross section distributions as a function of
  rapidity, $y_t$, of  the top and anti-top quark at LO (upper
  panel) and NLO (lower panel)  for  $ p\bar{p} \to t
  \bar{t} jj + X$ production  at the TeVatron run II with $\sqrt{s}=
  1.96 ~\textnormal{TeV}$.  The (orange) solid curve corresponds to the top
  quark, whereas the (brown) dashed  one to the anti-top quark.}
\end{figure}
\begin{figure}
\includegraphics[width=0.48\textwidth]{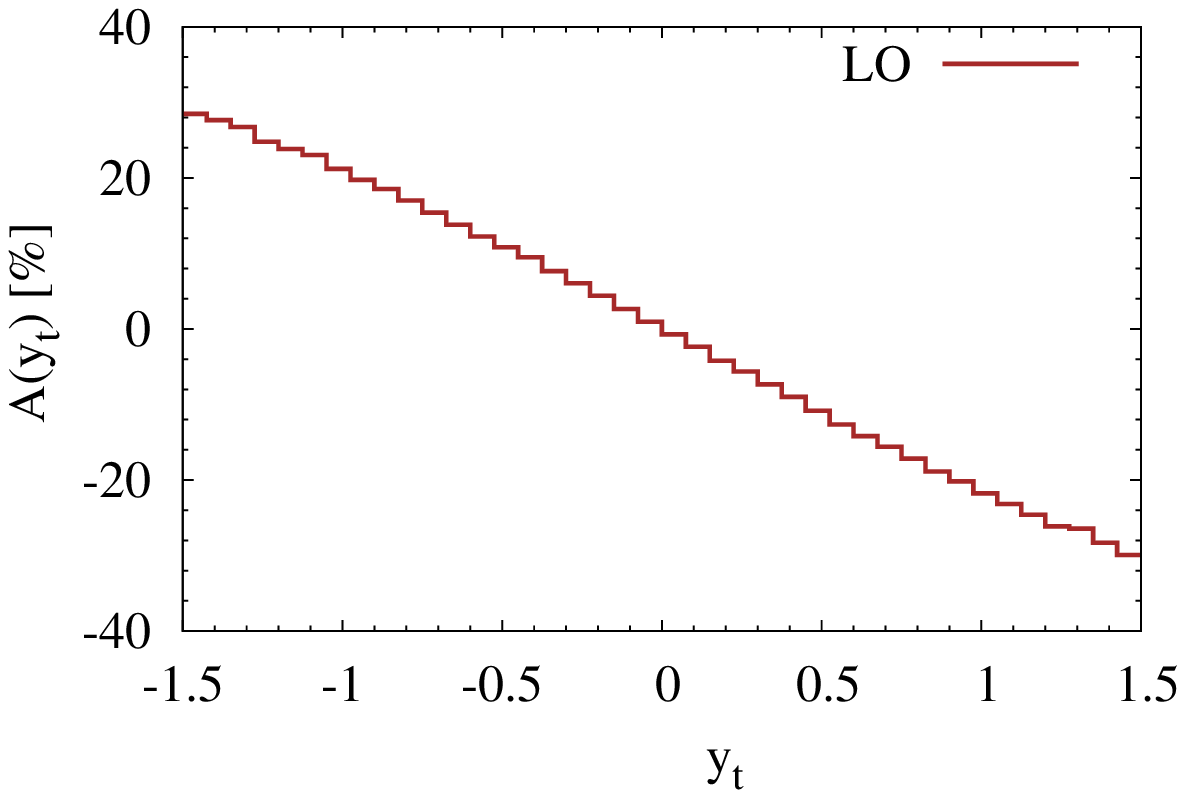}
\includegraphics[width=0.48\textwidth]{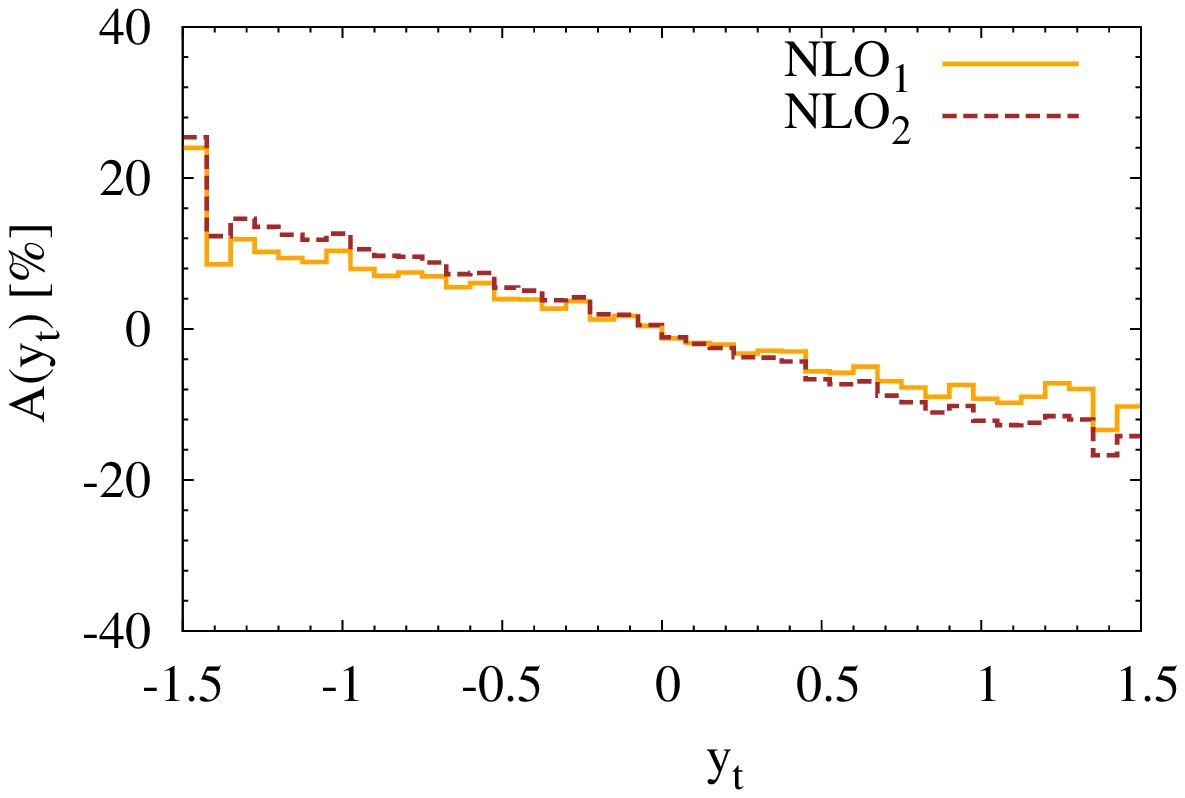}
\caption{\it \label{tev:asymmetry2} 
  Differential charge asymmetry, $A(y_t)$, as a function of the
  (anti-)top quark rapidity at LO (upper panel) and NLO
   (lower panel) for  $ p\bar{p} \to t \bar{t} jj + X$ production 
   at the TeVatron run II with $\sqrt{s}= 1.96 ~\textnormal{TeV}$. 
 $NLO_{1}$ refers to a result with a consistent expansion 
 in $\alpha_s$, while $NLO_{2}$ to the unexpanded one.}
\end{figure}
%
\begin{figure*}
\includegraphics[width=0.48\textwidth]{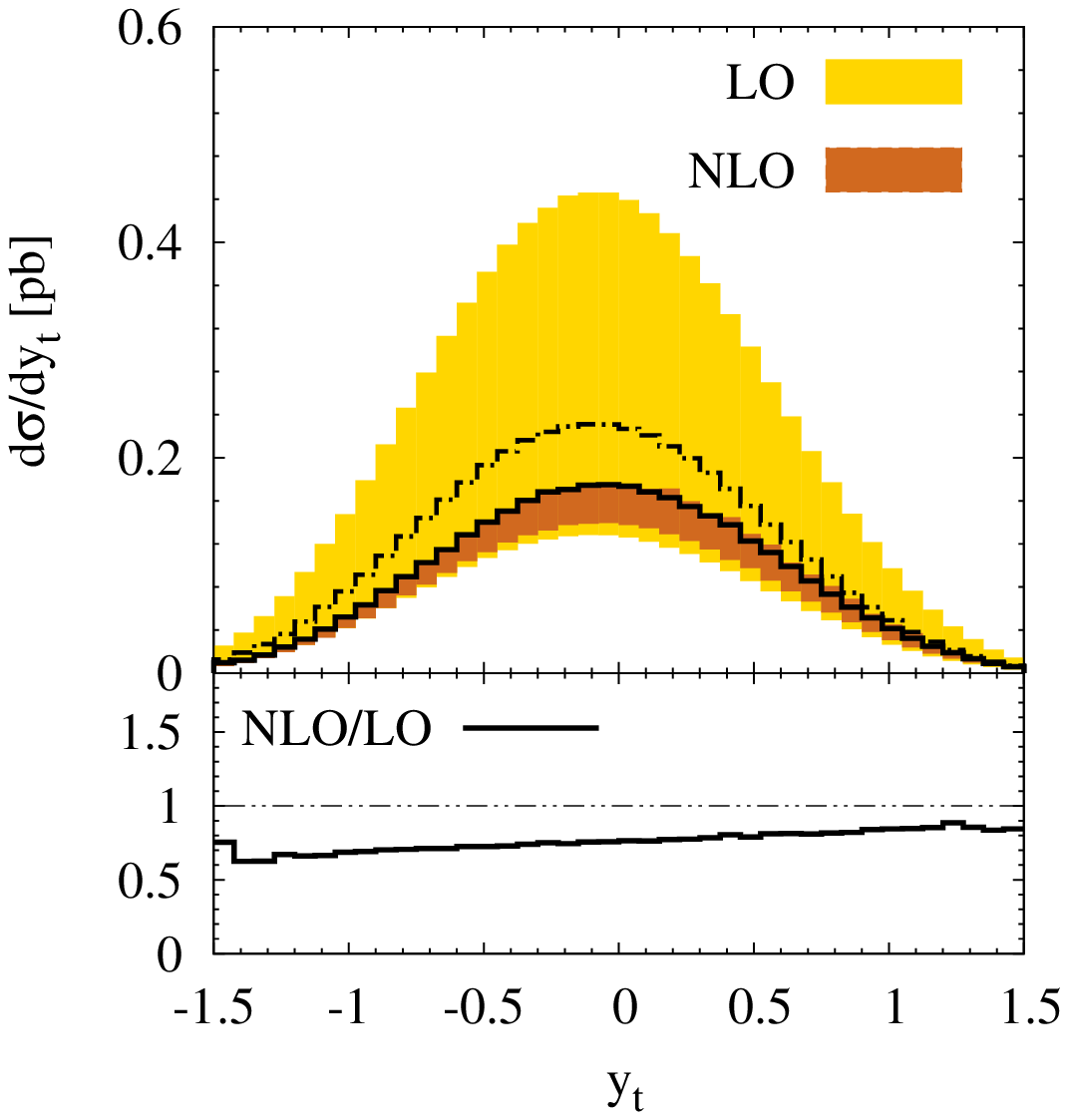}
\includegraphics[width=0.48\textwidth]{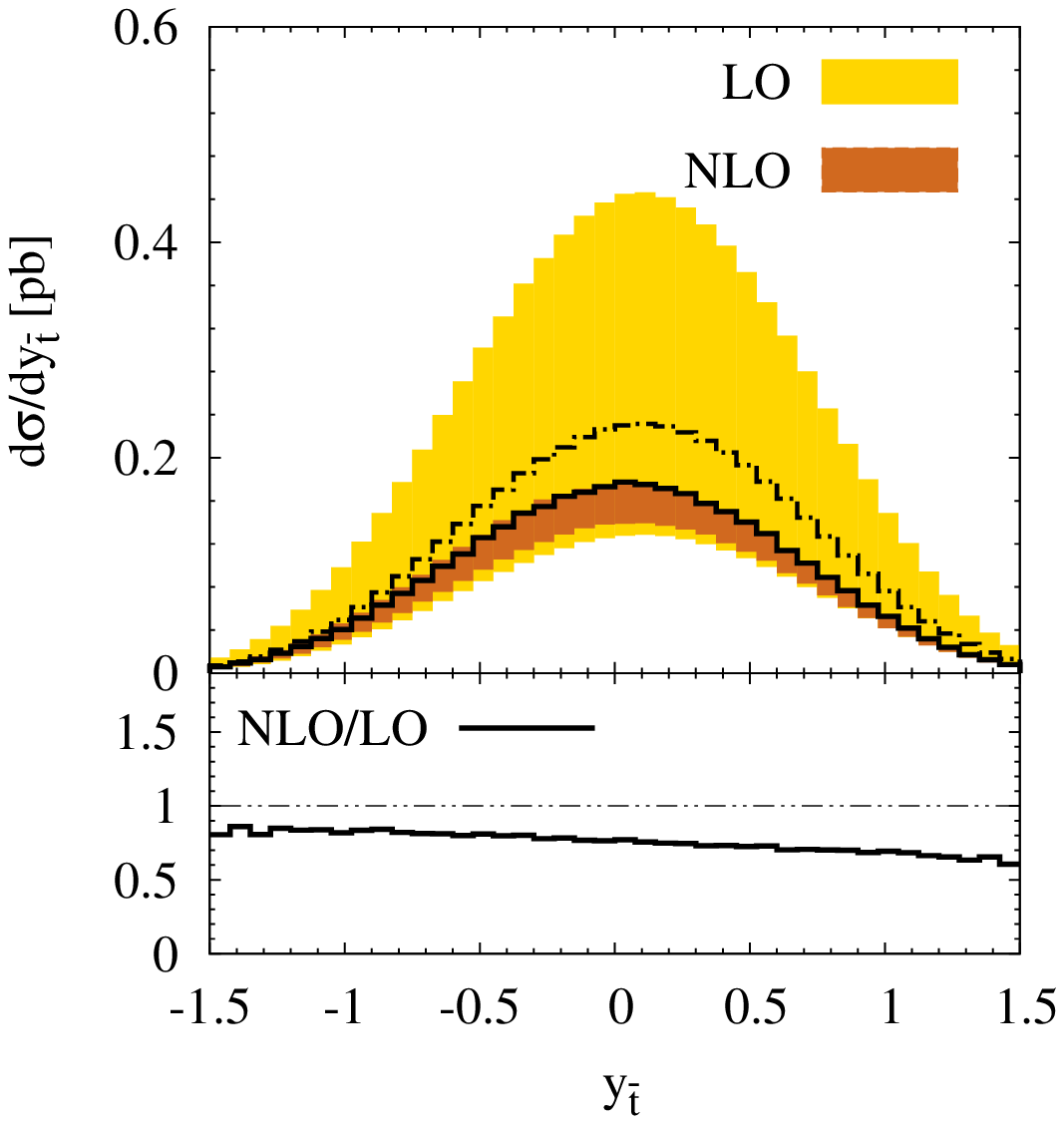}
\caption{\it \label{tev:fig1}  Differential cross section
  distributions as a function of the rapidity  of the top (left panel)
  and anti-top  (right panel) for  $ p\bar{p} \to t \bar{t} jj + X$
  production  at the TeVatron run II with $\sqrt{s}= 1.96
  ~\textnormal{TeV}$.  The dash-dotted curve corresponds to the LO,
  whereas the solid one to the NLO result. The uncertainty bands
  depict scale  variation. The lower panels display the differential
  $\cal K$ factor.}
\end{figure*}
\begin{figure}
\includegraphics[width=0.48\textwidth]{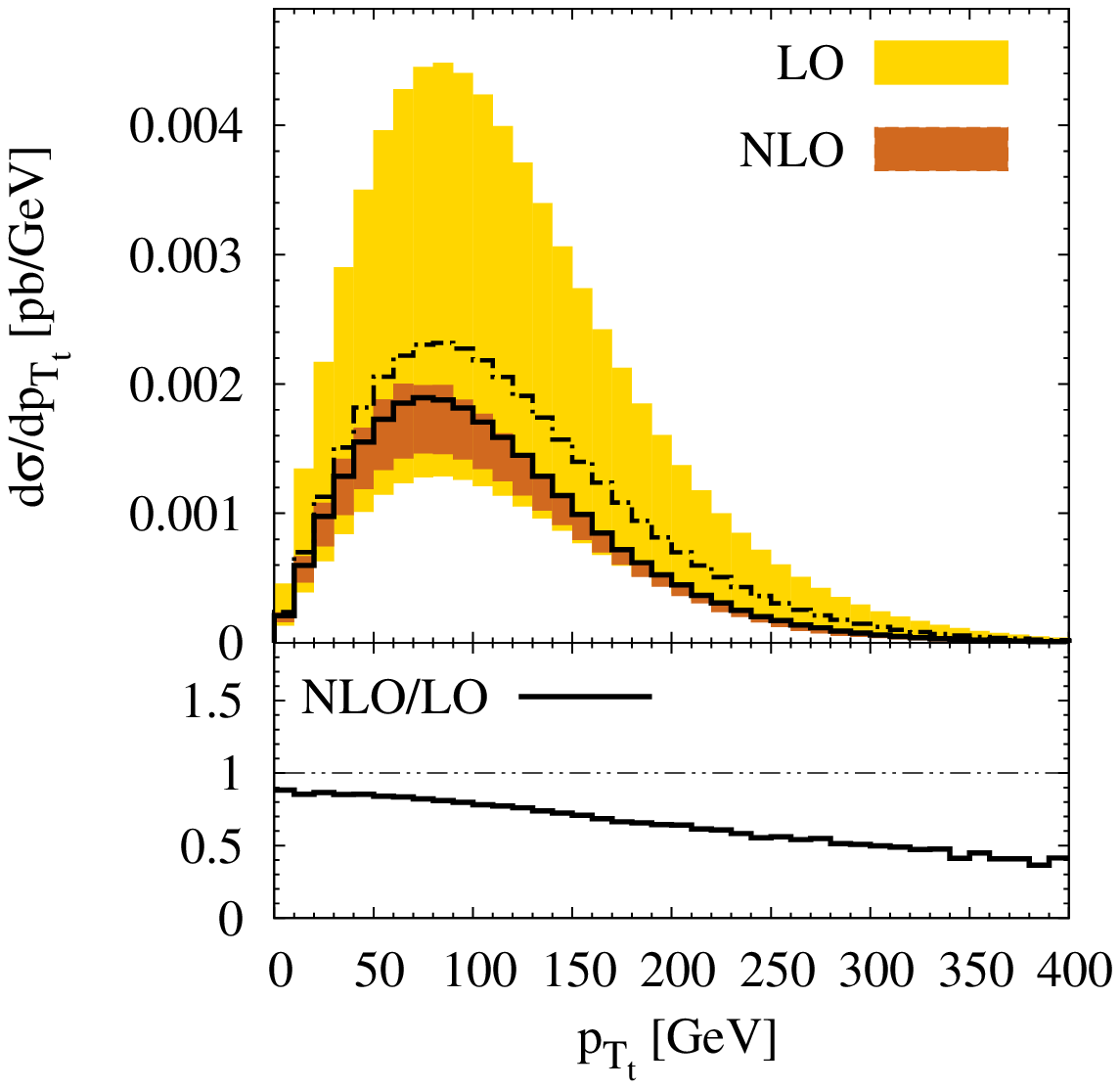}
\caption{\it \label{tev:fig2} Differential cross section distributions
  as a function of the averaged transverse momentum of the  (anti-)top
  for  $ p\bar{p} \to t \bar{t} jj + X$ production  at the TeVatron
  run II with $\sqrt{s}= 1.96 ~\textnormal{TeV}$.  The dash-dotted
  curve corresponds to the LO, whereas the solid one to the NLO
  result. The uncertainty bands depict scale  variation. The lower
  panels display the differential $\cal K$ factor.}
\end{figure}
\begin{figure*}
\includegraphics[width=0.48\textwidth]{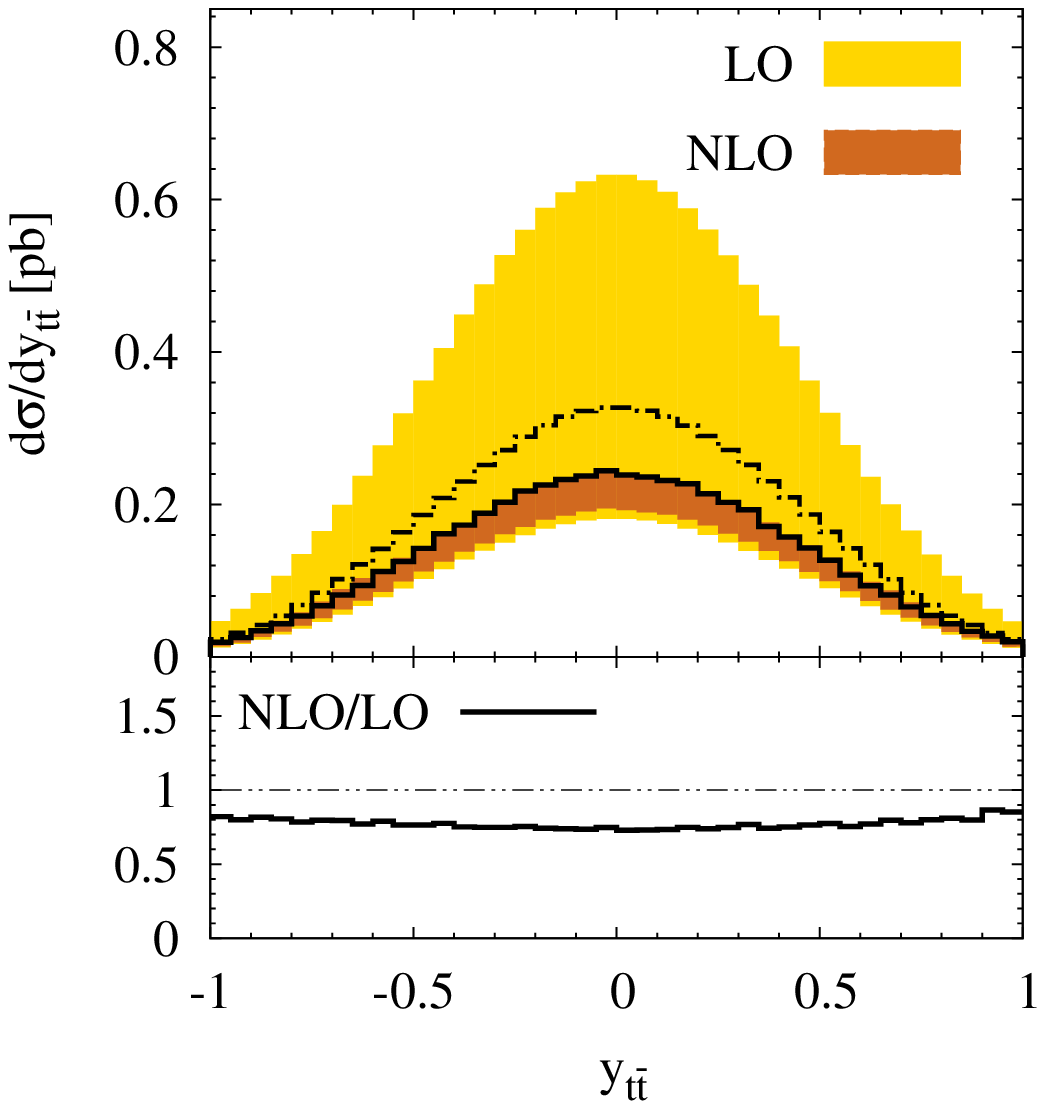}
\includegraphics[width=0.48\textwidth]{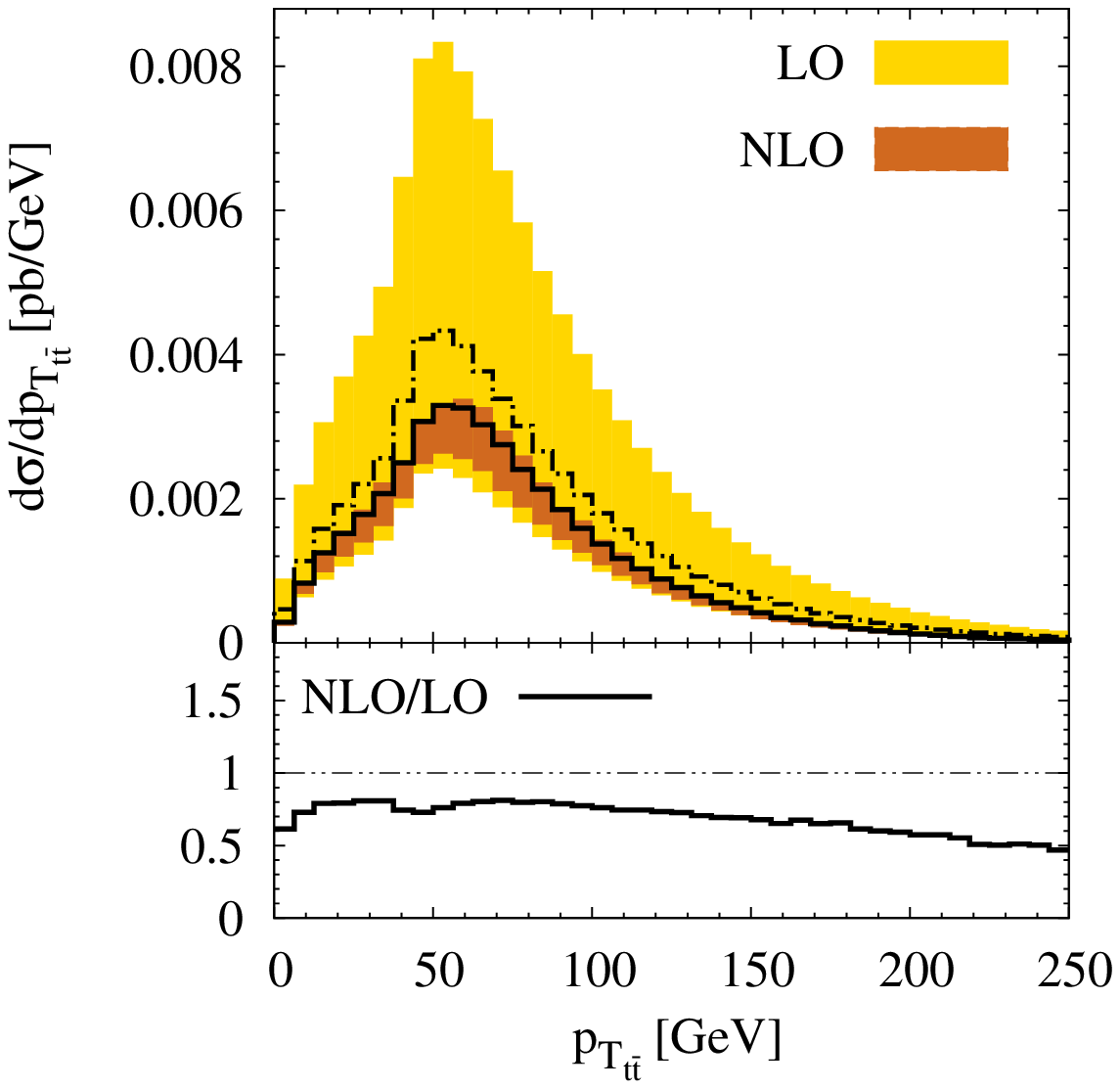}
\caption{\it  \label{tev:fig3} Differential cross section
  distributions as a function of rapidity (left panel) and  transverse
  momentum (right panel) of the  $t\bar{t}$ pair for  $ p\bar{p} \to t
  \bar{t} jj + X$ production  at the TeVatron run II with $\sqrt{s}=
  1.96 ~\textnormal{TeV}$.  The dash-dotted curve corresponds to the
  LO, whereas the solid one to the NLO result. The uncertainty bands
  depict scale variation. The lower panels display the differential
  $\cal K$ factor.}
\end{figure*}
\begin{figure*}
\includegraphics[width=0.48\textwidth]{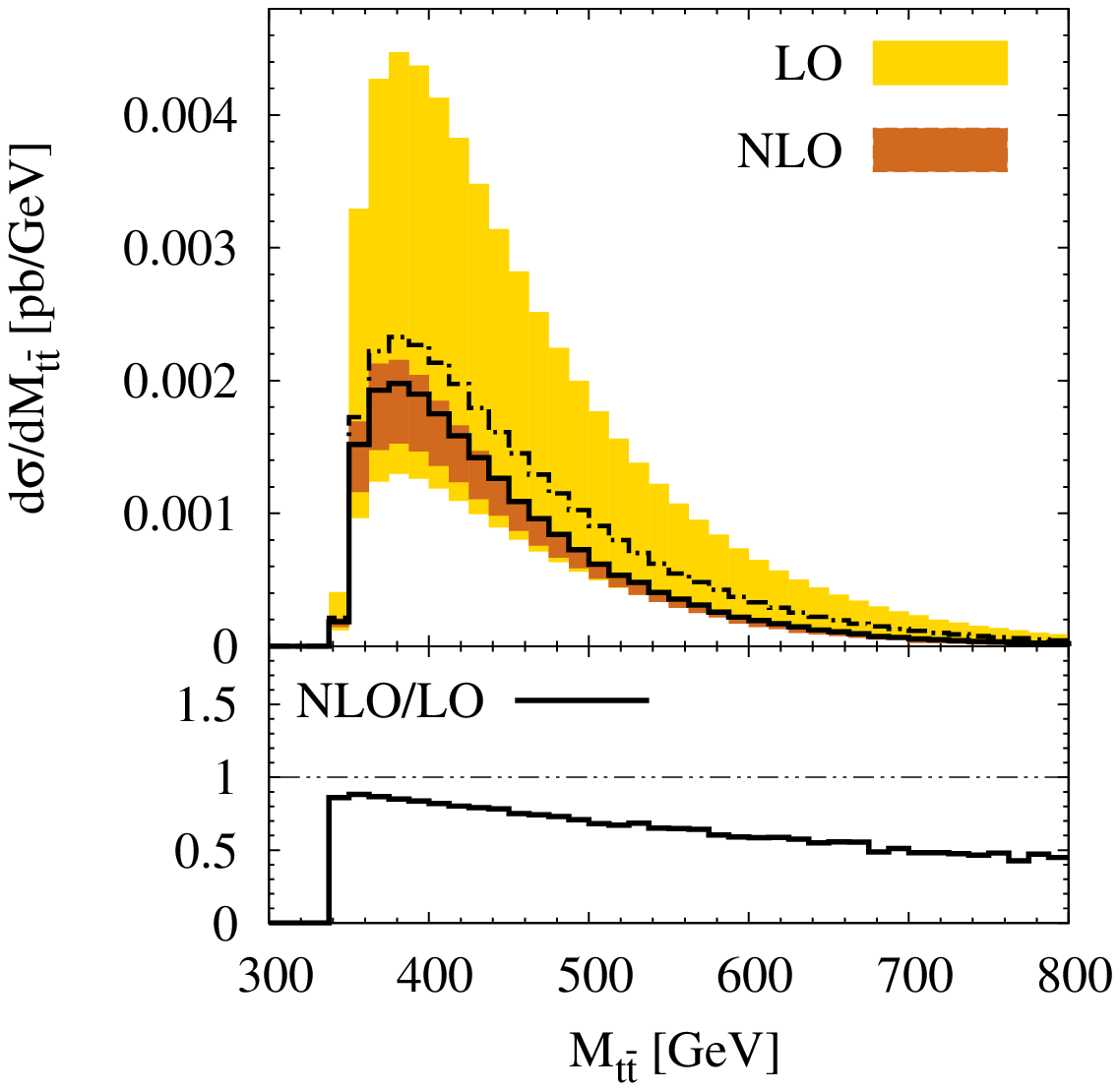}
\includegraphics[width=0.48\textwidth]{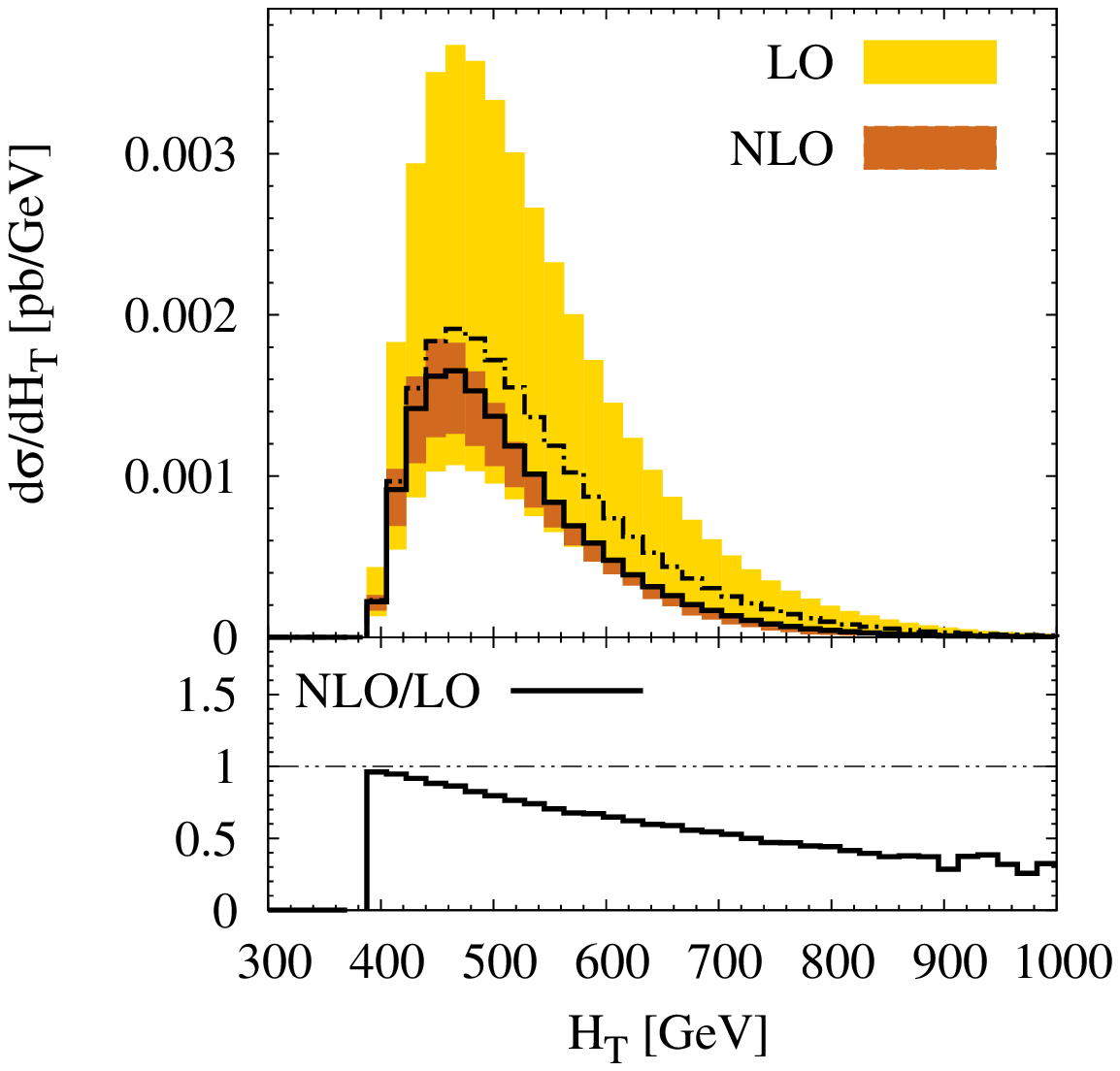}
\caption{\it  \label{tev:fig4} Differential cross section
  distributions as a function of the invariant mass of the  $t\bar{t}$
  pair (left panel) and of the total transverse energy (right panel)
  for  $ p\bar{p} \to t \bar{t} jj + X$ production  at the TeVatron
  run II with $\sqrt{s}= 1.96 ~\textnormal{TeV}$.  The dash-dotted
  curve corresponds to the LO, whereas the solid one to the NLO
  result. The uncertainty bands depict scale variation. The lower
  panels display the differential $\cal K$ factor.}
\end{figure*}
\begin{figure*}
\includegraphics[width=0.48\textwidth]{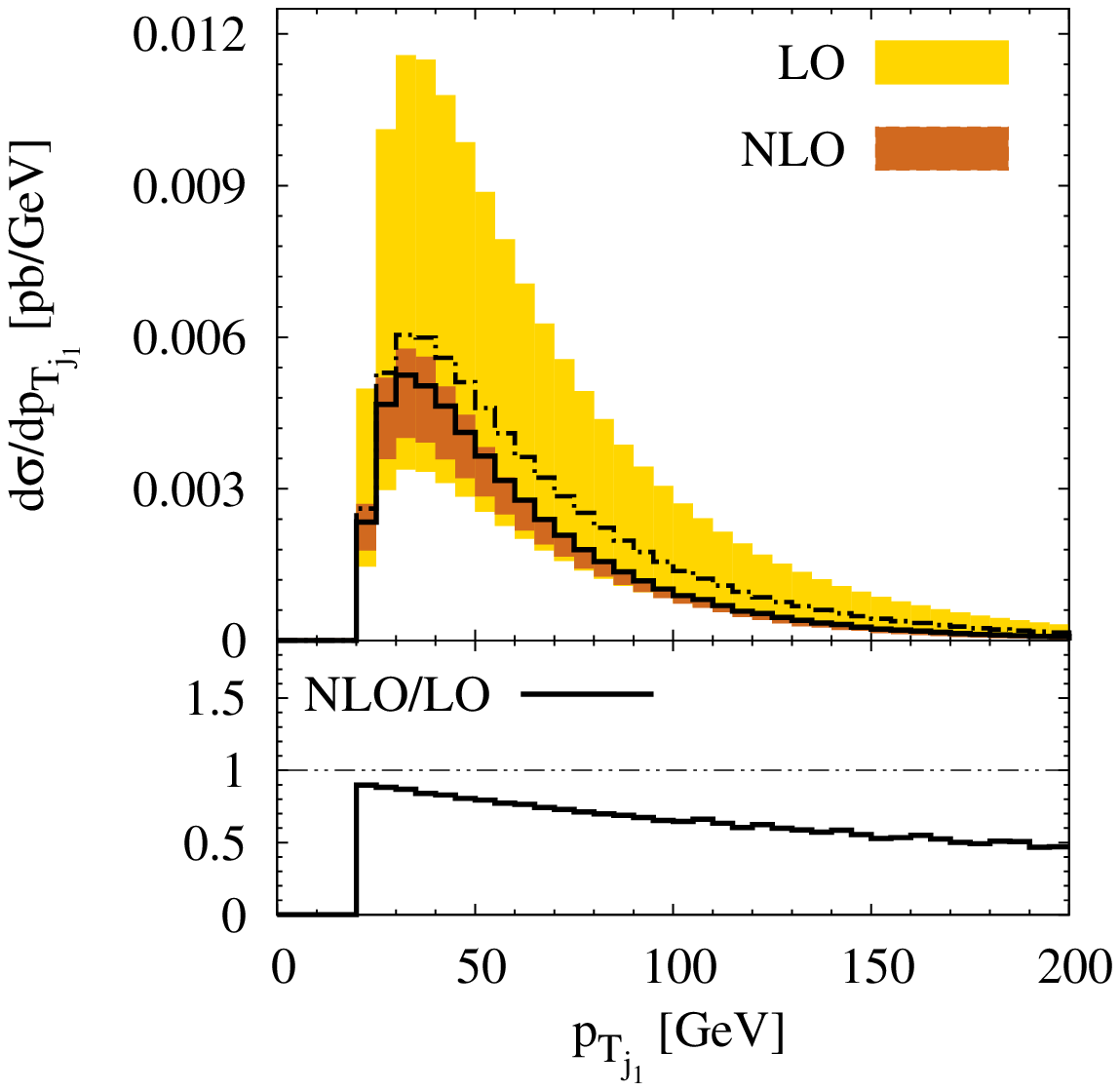}
\includegraphics[width=0.48\textwidth]{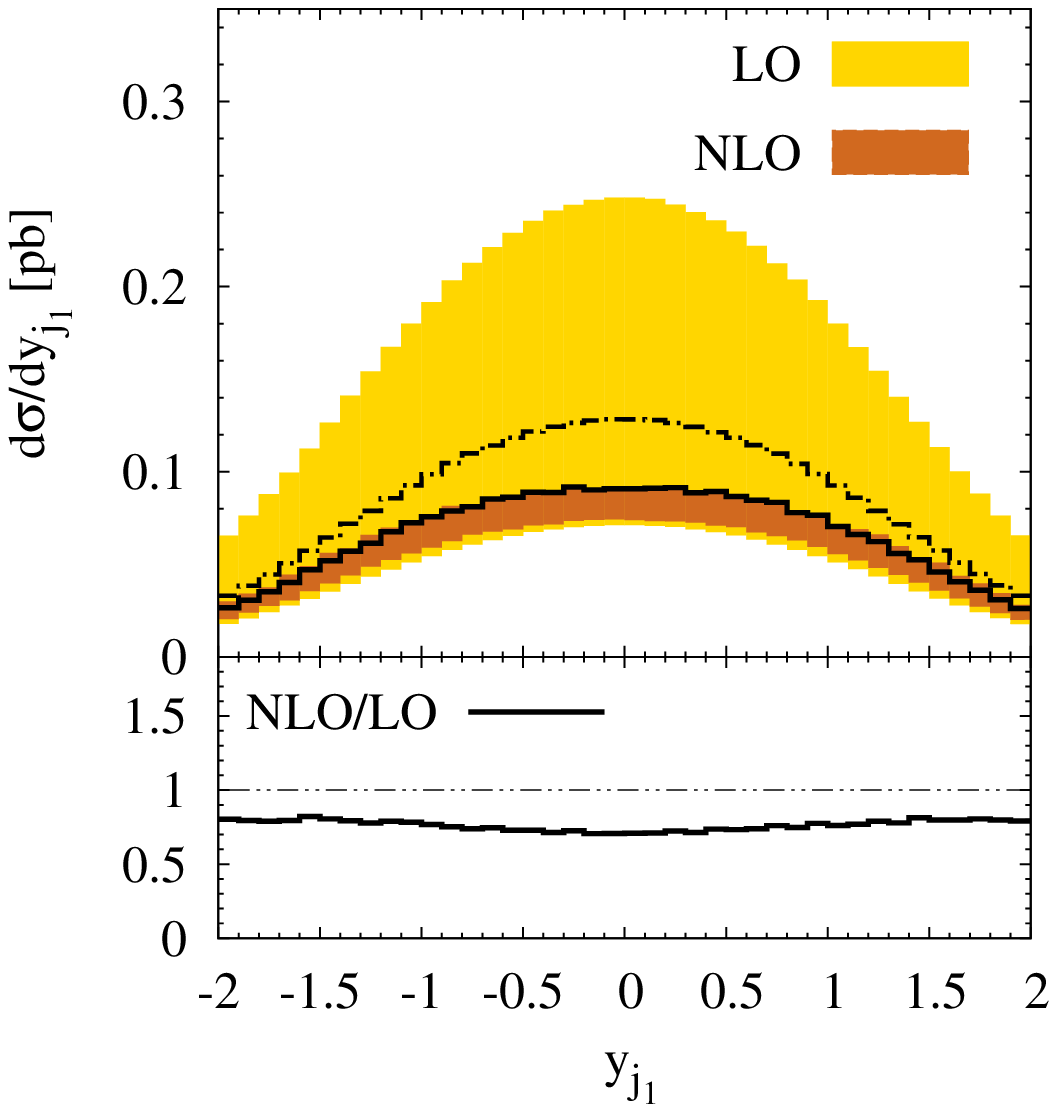}
\caption{\it  \label{tev:fig5}  Differential cross section
  distributions as a function of transverse momentum (left panel) and
  rapidity (right panel) of the 1st hardest jet for $ p\bar{p} \to t
  \bar{t} jj + X$ production at the TeVatron run II with $\sqrt{s}=
  1.96 ~\textnormal{TeV}$.  The dash-dotted curve corresponds to the
  LO, whereas the solid one to the NLO result. The uncertainty bands
  depict scale  variation. The lower panels display the differential
  $\cal K$ factor.}
\end{figure*}
\begin{figure*}
\includegraphics[width=0.48\textwidth]{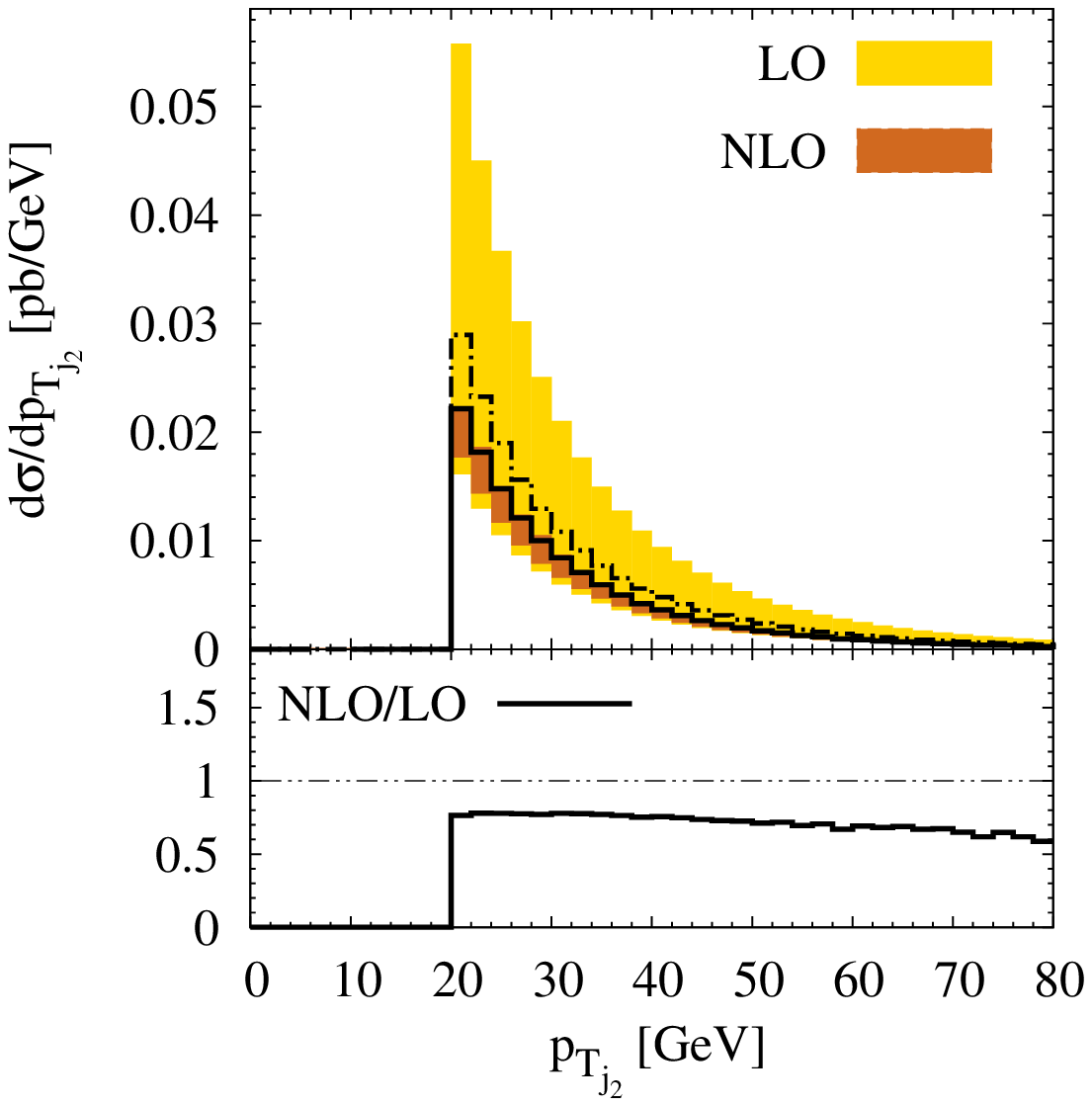}
\includegraphics[width=0.48\textwidth]{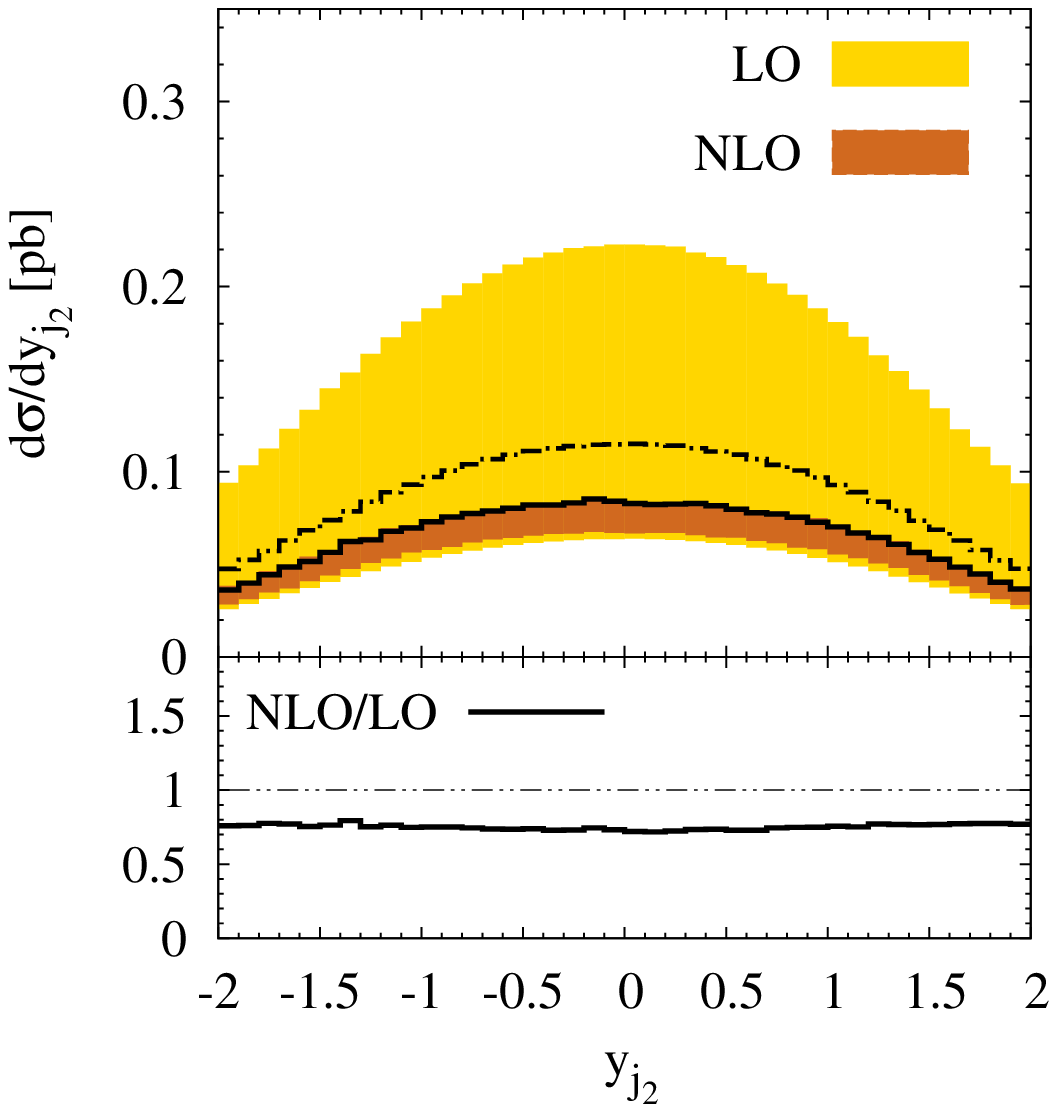}
\caption{\it  \label{tev:fig6}  Differential cross section
  distributions as a function of  transverse momentum (left panel) and
  rapidity (right panel) of the 2nd hardest jet  for $ p\bar{p} \to t
  \bar{t} jj + X$ production at the TeVatron run II with $\sqrt{s}=
  1.96 ~\textnormal{TeV}$.  The dash-dotted curve corresponds to the
  LO, whereas the solid one to the NLO result. The uncertainty bands
  depict scale  variation. The lower panels display the differential
  $\cal K$ factor.}
\end{figure*}
\begin{figure*}
\includegraphics[width=0.48\textwidth]{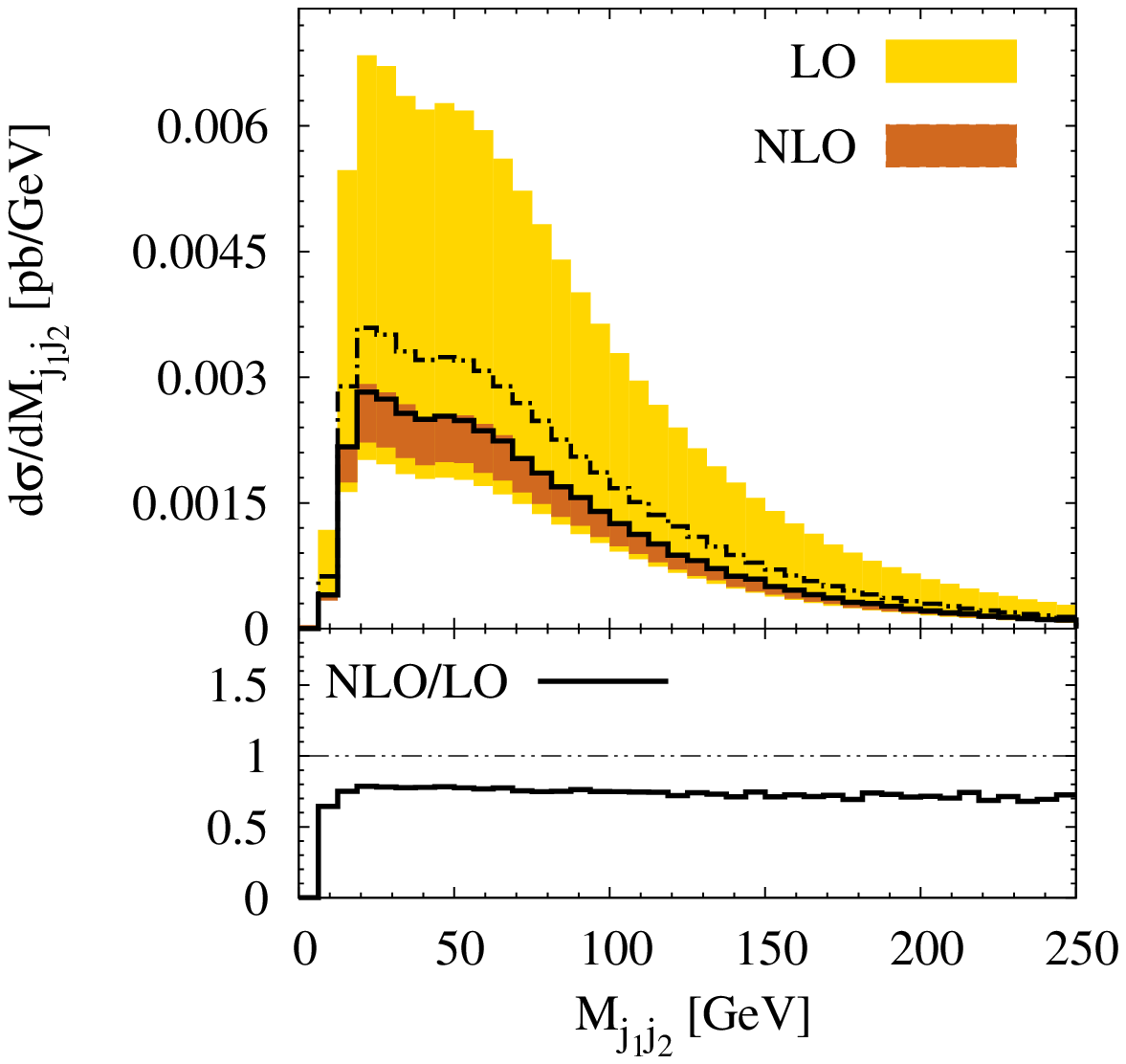}
\includegraphics[width=0.48\textwidth]{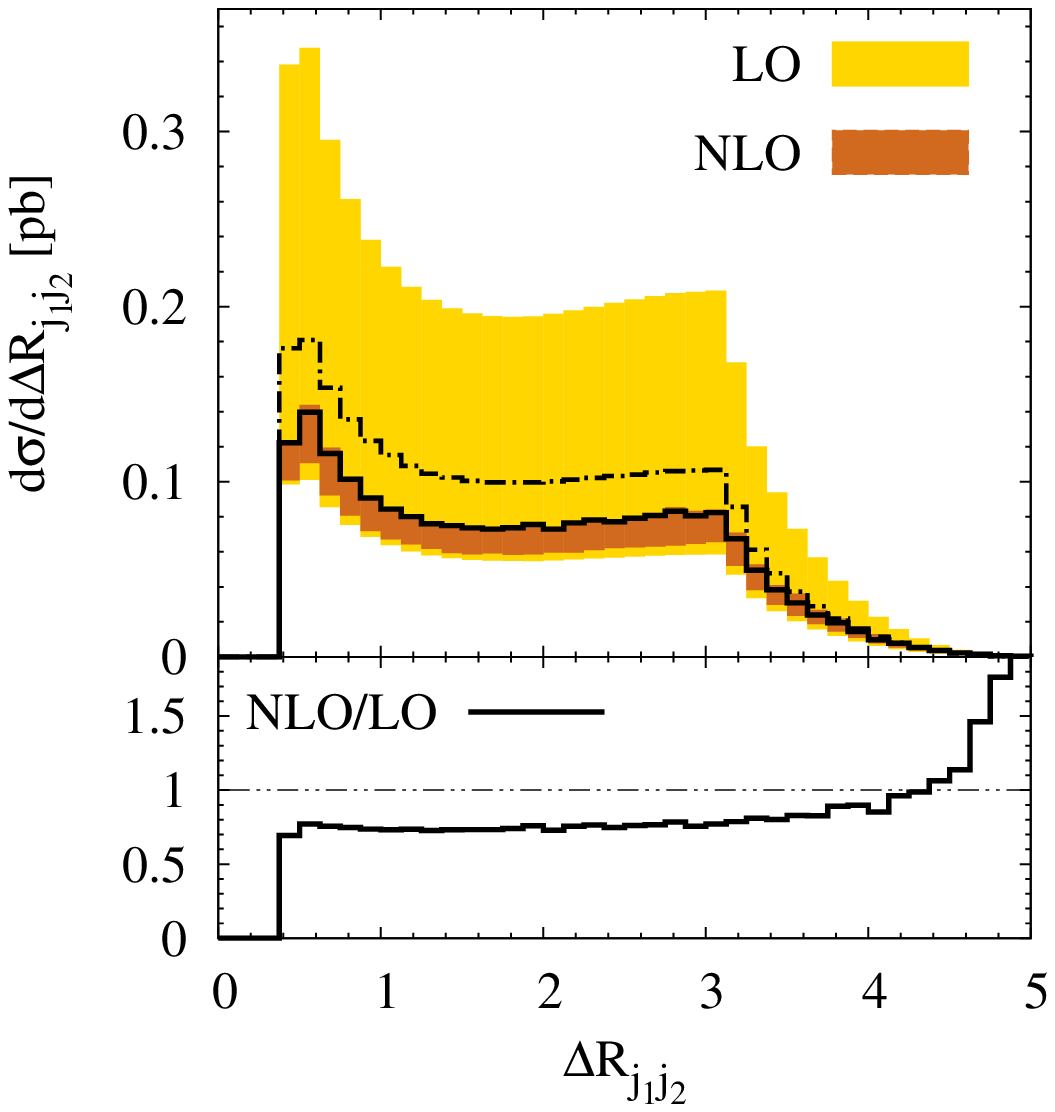}
\caption{\it  \label{tev:fig7}   Differential cross section
  distributions as a function of the invariant mass of the 1st and the
  2nd hardest jet (left panel) and $\Delta R_{jj}$ separation  (right
  panel) for $ p\bar{p} \to t \bar{t} jj + X$ production at the
  TeVatron run II with $\sqrt{s}= 1.96  ~\textnormal{TeV}$.  The
  dash-dotted curve corresponds to the LO, whereas the solid one to
  the NLO result. The uncertainty bands depict scale  variation. The
  lower panels display the differential $\cal K$ factor.}
\end{figure*}
%
Now we turn to a discussion of the numerical results.   We start with
the total
cross sections for  the TeVatron default selection. We find
\begin{equation}
\sigma_{\rm LO}^{\rm TeV}(pp\rightarrow t\bar{t}jj)
=0.3584^{\, +0.3390(94\%)}_{\, -0.1606(45\%)} 
~{\rm pb} \,,
\end{equation} 
\begin{equation}
\sigma_{\rm NLO}^{\rm TeV}(pp\rightarrow t\bar{t}jj)
= 0.2709^{\, +0.0014(0.5\%)}_{\, -0.0566( 21\%)} 
~{\rm pb} \,.
\end{equation}   
At the central value of the scale, $\mu_R=\mu_F=m_t$, the full cross
section receives moderate NLO correction of  the order of $-24\%$. The
scale dependence is indicated by  the upper and lower indices.  The
upper (lower) index represents the change when the scale is shifted
towards  $\mu=m_t/2$ ($\mu=2m_t$). Rescaling the common scale from the
default value $m_t$ up and down by a factor 2 changes the cross
section at LO by about  $94\%$. Through the inclusion of NLO QCD
corrections the scale uncertainty is dramatically reduced down to
$21\%$. 

To assess the effect of changing the jet algorithm, we compare NLO
results for the $k_T$,  {\it anti}$-k_T$  and the inclusive
Cambridge/Aachen (C/A) jet algorithms as presented in  Table
\ref{tab:tev1}.  With the jet resolution parameter of $R=0.4$ the
difference is within $1\%$.  
Total cross sections with a higher jets separation cut,
$\Delta R_{jj}>0.8$,  and the jet resolution  parameter $R=0.8$ are
shown in Table \ref{tab:tev2}. The NLO QCD corrections are  reduced
down  to $-16\%$ in that case. This is mostly due to a large, $-20\%$
change, in the  LO cross section as compared to a rather moderate,
$-9\%$ shift at NLO.  With an increase of the $R$ parameter the perturbative
difference between the algorithms should be more visible because the
phase  space for inclusion of an extra parton in the jet is much
larger. Nevertheless in our case it remains within $1\%$.  

For completeness   and as part of checks, the integrated NLO cross
sections for two different values of the $\alpha_{\rm max}$   parameter
are given in Table \ref{tab:tev3}. Let us remind the reader  that the
$\alpha_{\rm max}$  parameter is a common modification of subtraction
terms in the phase space away  from the singularity. We consider two
extreme choices, namely  $\alpha_{\rm max}=1$, which   corresponds to
the original formulation of \cite{Catani:2002hc},  and $\alpha_{\rm
  max}=0.01$. As can be observed, the independence of the final 
result on the value of $\alpha_{\rm max}$ is obtained at the permil level. 

In Table \ref{tab:tev4} we provide LO and NLO predictions  for the
integrated cross sections for  the TeVatron default selection,
however, for different values of the transverse  momentum cut for the
first  and the second hardest jet. The values are given for the
central scale  value $\mu_R=\mu_F=m_t$. Up to $\sim 50$ GeV, rather
moderate  corrections are acquired, however, for the high value of
$p_{T_{j}}$ cut, a substantial increase is noticed. For a
$p_{T_{j}}$ change by a factor of 4, NLO QCD corrections  are almost
doubled.   Compared to the total NLO $t\bar{t}$ cross section, 
\begin{equation}
\sigma_{\rm NLO}^{\rm TeV}(p\bar{p}\rightarrow t\bar{t})
=6.68^{\, +0.35(5.2\%)}_{\, -0.75(11.2\%)} 
~{\rm pb} \,,
\end{equation} 
obtained for the same setup, we find that
for the small $p_{T_j}$ cut  of  $20 ~{\rm GeV}$ the $t\bar{t}jj$
events represent only $4\%$ of  the total cross section. The fraction
is reduced substantially, down  to $0.8 \%$, $0.2\%$ and  $0.07\%$
respectively, if a $p_{T_j}$ cut  of $40 ~{\rm GeV}$, $60 ~{\rm GeV}$ 
and  $80 ~{\rm  GeV}$ is chosen instead. 
%
\subsubsection{Forward-Backward Asymmetry}
%
In the next step we calculate the top quark forward-backward asymmetry. At
LO the asymmetry is defined as 
\begin{equation}
{\cal A}_{\rm FB, LO}^{\rm t} =\frac{\sigma_{\rm LO}(y_t > 0)  - \sigma_{\rm
  LO}(y_t < 0)}{\sigma_{\rm LO}(y_t > 0)  + \sigma_{\rm
  LO}(y_t < 0)} \, ,
\end{equation}
where $\sigma_{\rm LO}^{\pm} = \sigma_{\rm LO}(y_t > 0)  \pm \sigma_{\rm
  LO}(y_t < 0)$ is evaluated with LO PDFs and LO $\alpha_s$. 
On the other hand, the asymmetry at NLO is defined by
\begin{equation}
\label{nloasymmetry}
{\cal A}_{\rm FB, NLO}^{\rm t}  = \frac{\sigma_{\rm LO}^{-}+\delta \sigma_{\rm
  NLO}^{-}}{\sigma_{\rm LO}^{+}+\delta \sigma_{\rm  NLO}^{+}} \, ,
\end{equation}
where $\delta \sigma_{\rm NLO}^{\pm}$ are the NLO contributions to the
cross sections and $\sigma_{\rm LO}^{\pm}$ are evaluated  with NLO
PDFs and NLO $\alpha_s$.  In case of large differences between the LO
and NLO asymmetries, it is necessary to expand
Eq. (\ref{nloasymmetry}) to first order in $\alpha_s$. Indeed, the
ratio in Eq. (\ref{nloasymmetry}) generates contributions of ${\cal
  O}(\alpha_s^2)$ and higher, which are affected by the unknown 
next-to-next-to-leading order contributions.  
The following definition has been used, for the same
reasons, in \cite{Dittmaier:2007wz,Dittmaier:2008uj}
\begin{equation}
{\cal A}_{\rm FB, NLO}^{\rm t}  = \frac{\sigma_{\rm LO}^{-}}{\sigma_{\rm LO}^{+}}
\left(  1 + \frac{\delta\sigma_{\rm NLO}^{-}}{\sigma_{\rm LO}^{-}} -
\frac{\delta \sigma_{\rm NLO}^{+}}{\sigma_{\rm LO}^{+}} \right) \, .
\end{equation}
The size of the forward-backward asymmetry of the top quark at LO  for 
$t\bar{t}jj$ production amounts to 
\begin{equation}
{\cal A}_{\rm FB, LO}^{\rm t}  = -0.103^{\,+0.003}_{\,-0.004}   \, .
\end{equation}
The NLO correction to the asymmetry as calculated  from 
Eq. (\ref{nloasymmetry}), and with a consistent expansion  in $\alpha_s$ reads
\begin{equation}\label{eq:nloexp}
{\cal A}_{\rm FB, NLO}^{\rm t}  = -0.046^{\,+0.005}_{\, -0.006} \, .
\end{equation}
Had we used an unexpanded ratio of the NLO cross sections, the result would 
rather be
\begin{equation}
{\cal A}_{\rm FB, NLO}^{\rm t}  = -0.058^{\, +0.014}_{\, -0.042} \, ,
\end{equation}
which means that the two definitions give reasonably consistent results for 
the central scale. However, the theoretical error
as calculated from the scale dependence is almost as large as the
asymmetry itself in the latter case.

In Figure \ref{tev:asymmetry1} rapidity distributions for the top and
anti-top quarks are presented  at LO and NLO. In both cases  results
are not symmetric around $y_t=0$ and are shifted  to a larger $y_t$
for the anti-top quarks and smaller $y_t$ for the top quarks.  This
shows that anti-top quarks are preferentially emitted in the
direction of the incoming protons (forward direction
by  definition), as implied by the sign in Eq.~\ref{eq:nloexp}. In
Figure \ref{tev:asymmetry2}, we have also plotted
the  differential charge asymmetry, ${\cal A}(y_t)$. In the NLO case
both results with and  without a consistent expansion in $\alpha_s$
are shown.  In both cases, it is clearly visible that the forward-backward
asymmetry of top quarks in $t\bar{t}jj$ production is significantly
reduced if NLO QCD corrections are taken into account.
%
\subsubsection{Differential cross sections}
%
In addition to the normalization of the integrated cross section,
radiative  corrections can affect the shape of various kinematic
distributions. To  quantify the size of these distortions we have
tested a variety  of  differential distributions.  

We start in Figure \ref{tev:fig1} with  the differential distribution
in  the rapidity of the top and anti-top, $y_t$, and in  Figure \ref{tev:fig2}
with an averaged  transverse momentum of the top and anti-top quarks.
The dash-dotted  curve corresponds to the LO,  whereas the solid one
to   the NLO result. The upper panels show the distributions themselves and
additionally include the scale-dependence bands obtained with a scale
variation by a factor of two. The lower  panels  display the ratio of the NLO value to 
the LO result, for the central scale values of $\mu=\mu_R=\mu_F=m_t$, called
the differential $\cal K$ factor. For the angular distributions we observe
negative corrections of the order of $15\%-30\%$.  More precisely,
$15\%$ corrections are observed  in the forward (backward) regions of
a  top (anti-top) distribution while $30\%$ corrections are in the
backward (forward) regions.   For the transverse momentum distribution
up to $100$ GeV moderate  and negative corrections of the order of $20\%$
are reached while the tail  exhibits even  $-60 \%$ corrections.  

 In Figure \ref{tev:fig3}  and Figure \ref{tev:fig4}  differential
 cross section distributions as a function of rapidity,
 $y_{t\bar{t}}$, transverse momentum, $p_{T_{t\bar{t}}}$, and
 invariant mass, $m_{t\bar{t}}$, of the $t\bar{t}$ pair are presented. 
 Also shown is the  differential distribution in the total transverse energy 
 of the system, $H_T$. The latter is  defined as 
\begin{equation}
H_T = m_{T_t} + m_{T_{\bar t}}+ \sum_{i=1,2} p_{T_{j_i}}  \, ,
\end{equation}
where
 \begin{equation}
m_{T_t}=\sqrt{m^2_t+p_{T_t}^2}\, .
 \end{equation}
The rapidity distribution of the $t\bar{t}$ pair has almost  constant,
negative and moderate    $20\%-25\%$ corrections.  Both,
$m_{t\bar{t}}$ and $p_{T_{t\bar{t}}}$ distributions get large negative
$50\%-60\%$ corrections for the higher values of these observables,
which have to be compared with also negative, but $20\%-25\%$
corrections close  to the $t\bar{t}$ threshold for $m_{t\bar{t}}$ and
for smaller values of the transverse momentum of the $t\bar{t}$ pair
for the $p_{T_{t\bar{t}}}$ distribution. At the
begin of the $H_T$ spectrum  negative and small $10\%-15\%$
corrections can be seen. However, at the same  time even $-70\%$ can
be reached for higher values. 

In the next step, the jet kinematics is presented. We start with
transverse momentum and rapidity distributions of  the $1^{\rm st}$
and the $2^{\rm nd}$ hardest jet as depicted in  Figure \ref{tev:fig5}
and Figure \ref{tev:fig6}. Again, in case of rapidity distributions,
negative and almost constant  $20\%-25\%$  corrections are obtained. 
On the other hand, also negative $40\%-50\%$ corrections are visibled at  
the tails of $p_{T_ {j}}$ spectrums, as compared to minus $20\%$ in the low
transverse momentum range. 

And finally, in Figure  \ref{tev:fig7},  the invariant mass, $m_{jj}$, of the
$1^{\rm st}$ and the $2^{\rm nd}$ hardest jet together with their separation
in the $(y,\phi)$ plane are shown. In both cases almost constant and rather
moderate differential ${\cal K}$-factors of $0.75$ can be seen. Except for the
tails of $\Delta R_{jj}$ distribution where distortions are much
higher. However, fluctuations visible there are a reflection  of the limited
statistics of the Monte Carlo integration.  At the TeVatron, $\Delta R_{jj}$ 
is dominated by the transverse plane peak at the begin of the spectrum, which 
suggests that two hardest jets are preferably produced almost collinear to each
other.

Overall, we can say that at the TeVatron employing a fixed scale
$\mu=\mu_R=\mu_F=m_t$, the NLO corrections to transverse momentum
distributions are rather substantial.  Moreover, they do not simply
rescale the LO shapes, but induce distortions at the level of
$40\%$. The  same applies to the invariant mass distribution of the
$t\bar{t}$ pair. In case of the total transverse energy, even $50\%-60\%$
deformations can be observed. For angular distributions and the
invariant mass of two tagging jets we observe negative and rather
moderate corrections of the order of $15\%-30\%$, which turn out to be
relatively constant.
%
\subsection{LHC}
%
As in the TeVatron case we calculate the partonic cross sections for events
with at least two hard  jets. Jets are obtained via the {\it anti}$-k_T$ jet
algorithm  with a resolution parameter $R=0.5$ and are required to have
\begin{equation}
p_{T_{j}} > 50 ~{\rm GeV}\, , ~~~|y_j| < 2.5\, , ~~~\Delta R_{jj} >
0.5\, .
\end{equation}
We will refer to this set of cuts as {\it LHC default selection}.
All results presented in this Section are given in picobarns.
%
\subsubsection{Integrated cross sections}
%
\begin{table}
  \caption{\it \label{tab:lhc1} Integrated cross sections at LO and NLO for
    $pp\rightarrow t\bar{t}jj ~+ X$  production  at the LHC with
    $\sqrt{s}= 7 ~\textnormal{TeV}$.  Results for three different jet
    algorithms are presented,   the anti-$k_T$, $k_T$ and the Cambridge/Aachen
    jet algorithms with $R=0.5$. The scale choice is
    $\mu_R=\mu_F=m_{t}$. }
\begin{ruledtabular}
  \begin{tabular}{lcccr}
  \textsc{Cuts}  &  
  $\sigma_{\rm LO}$ [pb]       &
  $\sigma_{\rm NLO}^{\rm anti-k_T}$  [pb] &   
  $\sigma_{\rm NLO}^{\rm k_T}$   [pb] &   
  $\sigma_{\rm NLO}^{\rm C/A}$   [pb]\\
&&&&\\
\hline
   $p_{T_{j}} > 50$ GeV  &&&&\\
   $\Delta R_{jj}> 0.5$ & 13.398(4) & 9.82(2) & 9.86(2) & 9.86(2) \\
  $|y_j| < 2.5$ && &&\\
  \end{tabular}
\end{ruledtabular}
\end{table}
\begin{table}
  \caption{\it \label{tab:lhc2} Integrated cross section at LO and NLO for 
    $pp\rightarrow t\bar{t}jj ~+ X$  production  at the LHC with
    $\sqrt{s}= 7 ~\textnormal{TeV}$. Results   for three different jet
    algorithms are presented,  the anti-$k_T$, $k_T$ and the Cambridge/Aachen
    jet algorithms with $R=1.0$. The scale choice is $\mu_R=\mu_F=m_{t}$.}
\begin{ruledtabular}
  \begin{tabular}{lcccr}
     \textsc{Cuts}  &
     $\sigma_{\rm LO}$ [pb]       &
     $\sigma_{\rm NLO}^{\rm anti-k_T}$  [pb] &   
     $\sigma_{\rm NLO}^{\rm k_T}$   [pb] &   
     $\sigma_{\rm NLO}^{\rm C/A}$   [pb] \\
 &&&&\\
\hline
      $p_{T_{j}} > 50$ GeV &&&&  \\
     $\Delta R_{jj}> 1.0$ &11.561(4) & 9.95(2)  & 10.06(2) &  10.04(2) \\
     $|y_j| < 2.5$  &&&&  \\
  \end{tabular}
\end{ruledtabular}
\end{table}
\begin{table}
  \caption{\it \label{tab:lhc3} Integrated cross section at  NLO for 
    $pp\rightarrow t\bar{t}jj ~+ X$  production  at the LHC
    with $\sqrt{s}= 7  ~\textnormal{TeV}$.  Results for two
    different values of the $\alpha_{\rm max}$  parameter and $\Delta R_{jj}$ cut
    are presented. They are obtained    with the anti-$k_T$ jet algorithm. 
    The scale choice is $\mu_R=\mu_F=m_{t}$.}
\begin{ruledtabular}
  \begin{tabular}{lcr}
   \textsc{Cuts}    & 
   $\sigma_{\rm NLO}^{\rm \alpha_{max}=0.01}$ [pb]& 
   $\sigma_{\rm NLO}^{\rm \alpha_{max}=1.00}$   [pb]    \\
 &&\\
\hline
   $\Delta R_{jj}>0.5$, $R=0.5$  &   9.82(2)  &   9.81(1) \\
   $\Delta R_{jj}>1.0$, $R=1.0$  &   9.95(2)  &   9.94(1)   \\
  \end{tabular}
\end{ruledtabular}
\end{table}
\begin{table}
  \caption{\it \label{tab:lhc4} Integrated cross section at LO and NLO for 
    $pp\rightarrow t\bar{t}jj ~+ X$  production  at the LHC with
    $\sqrt{s}= 7 ~\textnormal{TeV}$. Results   for the anti-$k_T$ jet
    algorithm with $R=0.5$ are presented.  In the last two columns the $\cal
    K$ factor, defined as  the ratio of the NLO cross section to the
    respective LO result, and NLO corrections in \% are given. The scale
    choice is $\mu_R=\mu_F=m_{t}$.}
\begin{ruledtabular}
  \begin{tabular}{lcccr}
     \textsc{$p_{T_j}$ Cut}  &
     $\sigma_{\rm LO}$ [pb]       &
     $\sigma_{\rm NLO}^{\rm anti-k_T}$  [pb] & $\cal{K}$ & [\%] \\
&&&&\\
\hline
     $p_{T_{j}} > 50$ GeV & 13.398(4) & 9.82(2)  & 0.73  & -27 \\
     $p_{T_{j}} > 75$ GeV & 5.944(2) & 4.115(8) & 0.69  & -31 \\
     $p_{T_{j}} > 100$ GeV  & 3.018(1) & 1.944(4) & 0.64  & -36\\
     $p_{T_{j}} > 125$ GeV & 1.665(1) & 0.993(2) & 0.60  & -40 \\
  \end{tabular}
\end{ruledtabular}
\end{table}
%
\begin{figure*}
\includegraphics[width=0.48\textwidth]{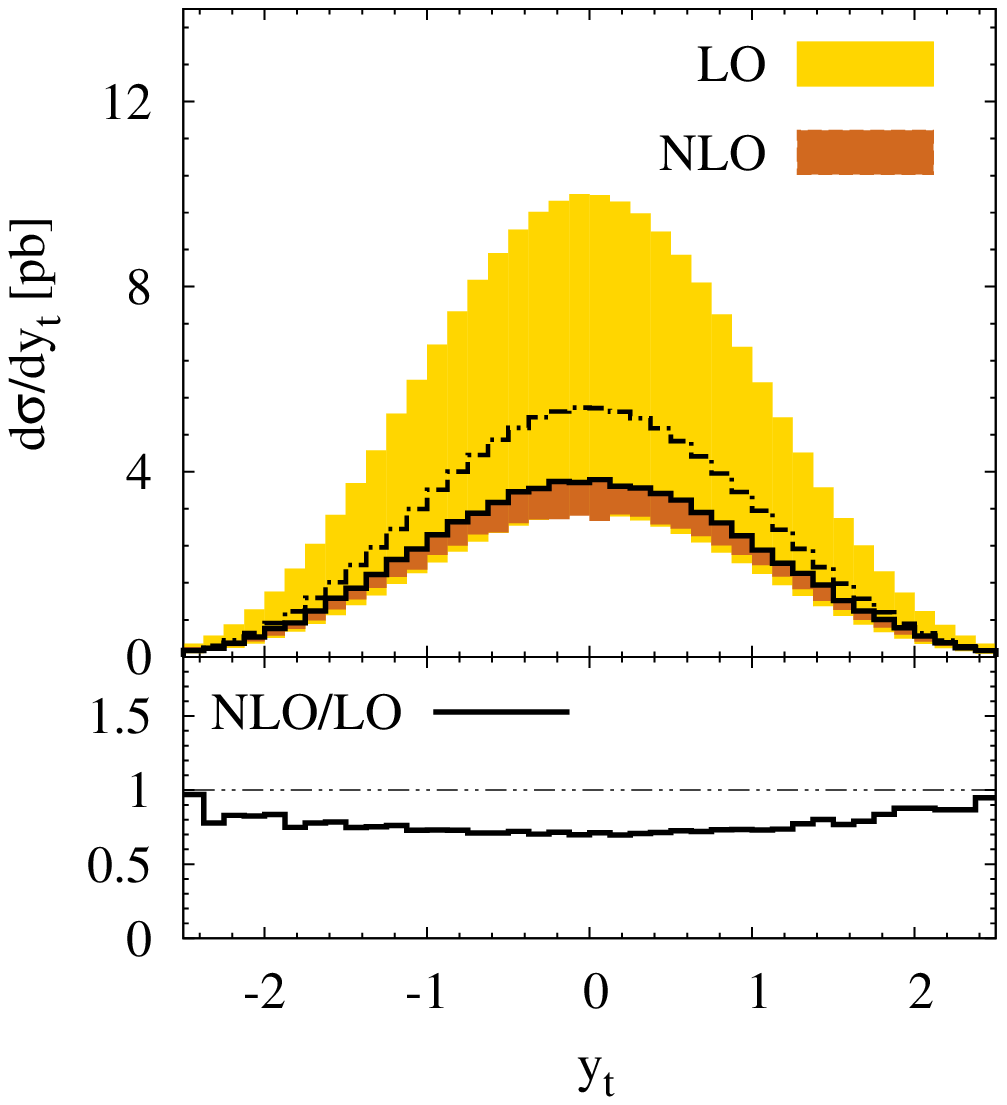}
\includegraphics[width=0.48\textwidth]{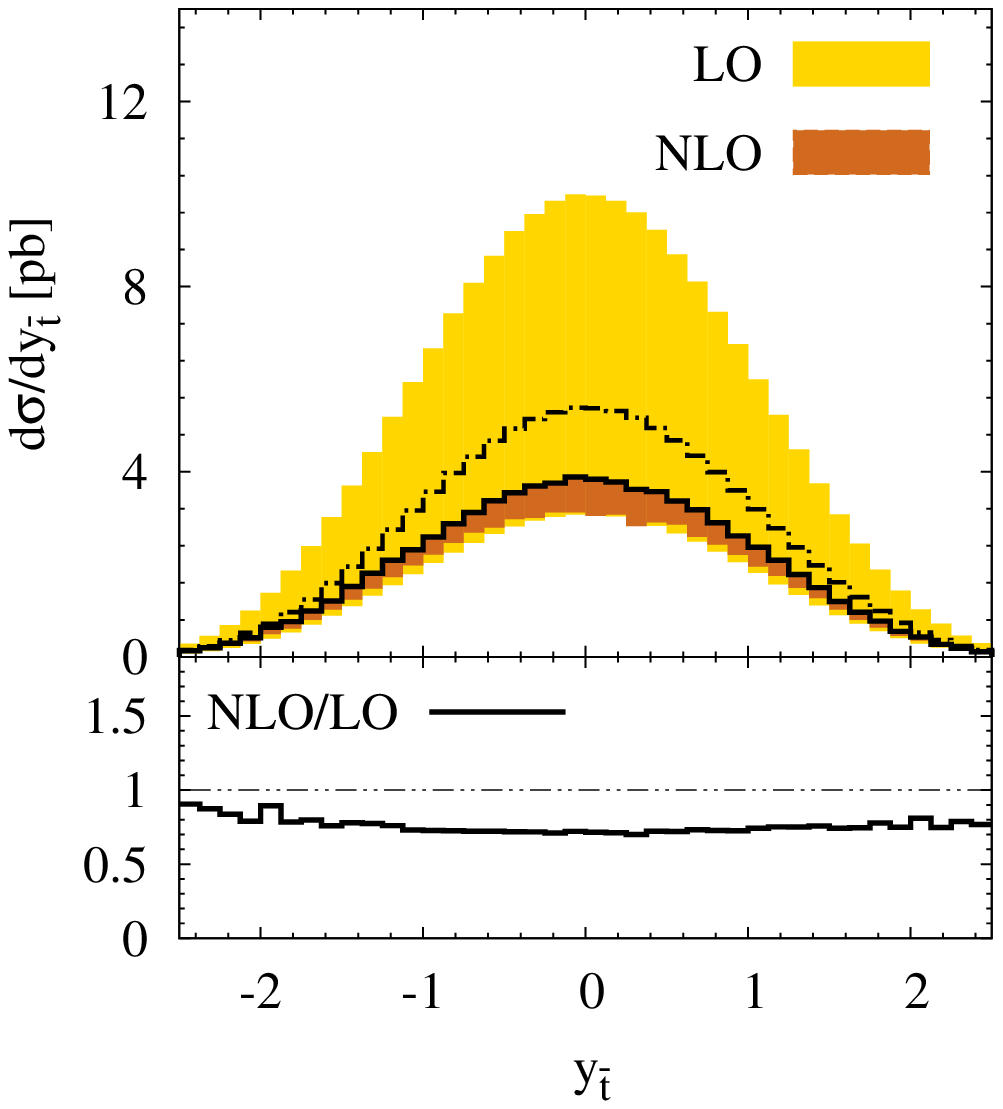}
\caption{\it \label{fig:lhc1} Differential cross section distributions
  as a function of the rapidity of the top quark (left panel) and of
  the anti-top quark (right panel) for  $ pp \to t \bar{t} jj + X$
  production at the LHC with $\sqrt{s}= 7 ~\textnormal{TeV}$. The
  dash-dotted curve corresponds to the LO, whereas the solid one to
  the NLO result. The uncertainty bands depict scale  variation.  The
  lower panels display the differential $\cal K$ factor. }
 \end{figure*}
\begin{figure}
\includegraphics[width=0.48\textwidth]{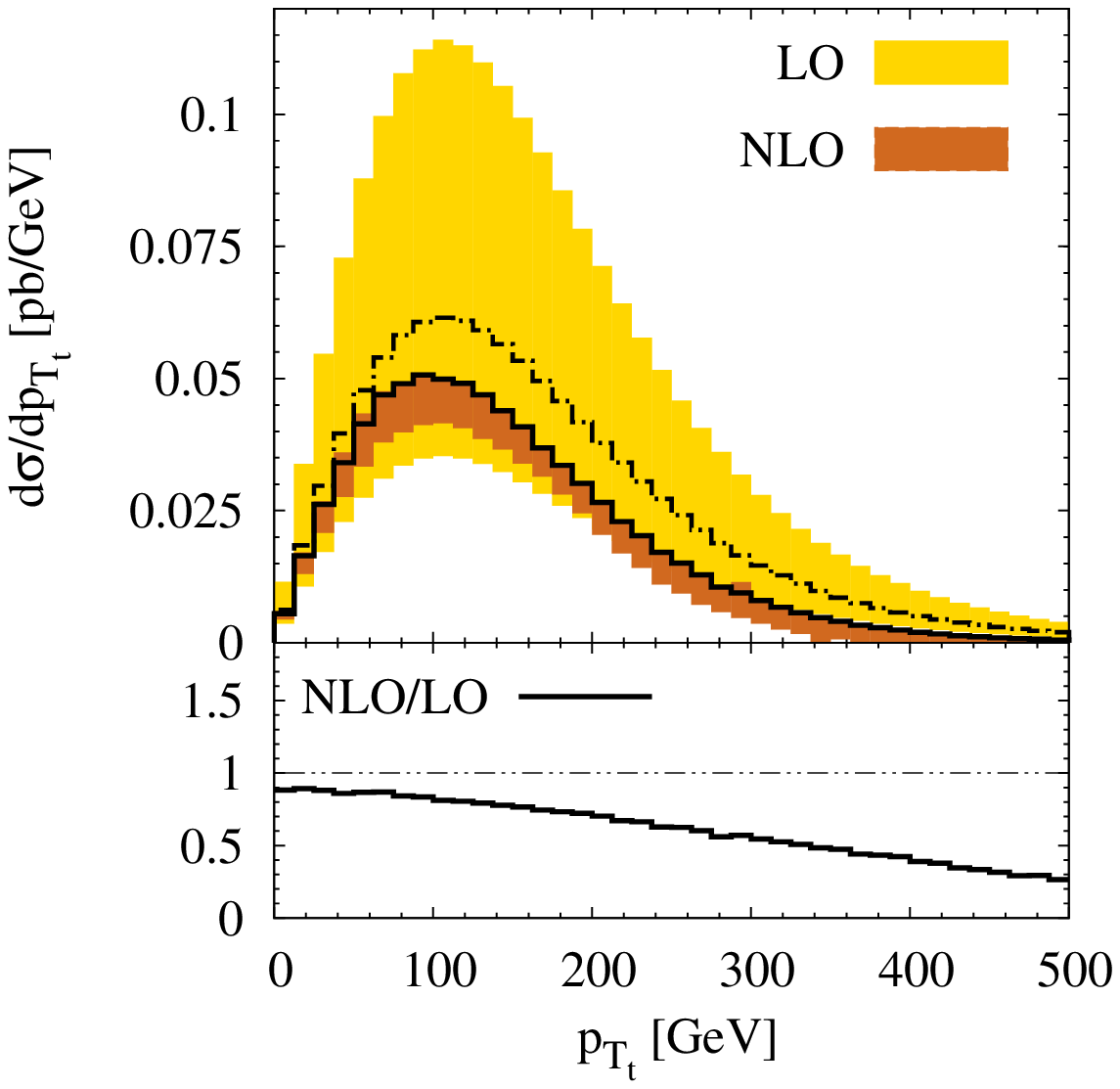}
\caption{\it \label{fig:lhc2} Differential cross section distributions
  as a function of the averaged transverse momentum of the top and
  anti-top for  $ pp \to t \bar{t} jj + X$ production at the LHC with
  $\sqrt{s}= 7 ~\textnormal{TeV}$. The dash-dotted curve corresponds
  to the LO, whereas the solid one to the NLO result. The uncertainty
  bands depict scale  variation.  The lower panels display the
  differential $\cal K$ factor. }
 \end{figure}
\begin{figure*}
\includegraphics[width=0.48\textwidth]{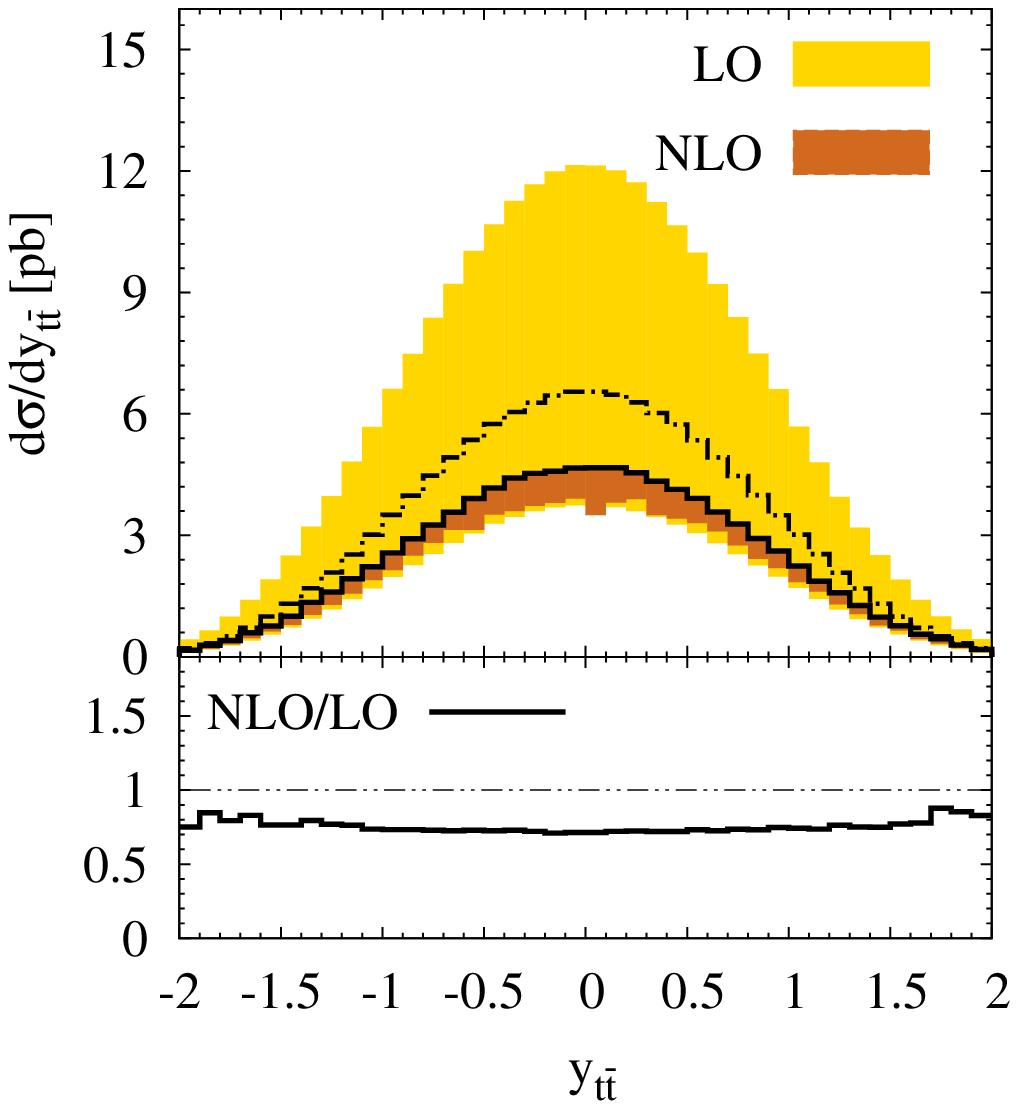}
\includegraphics[width=0.48\textwidth]{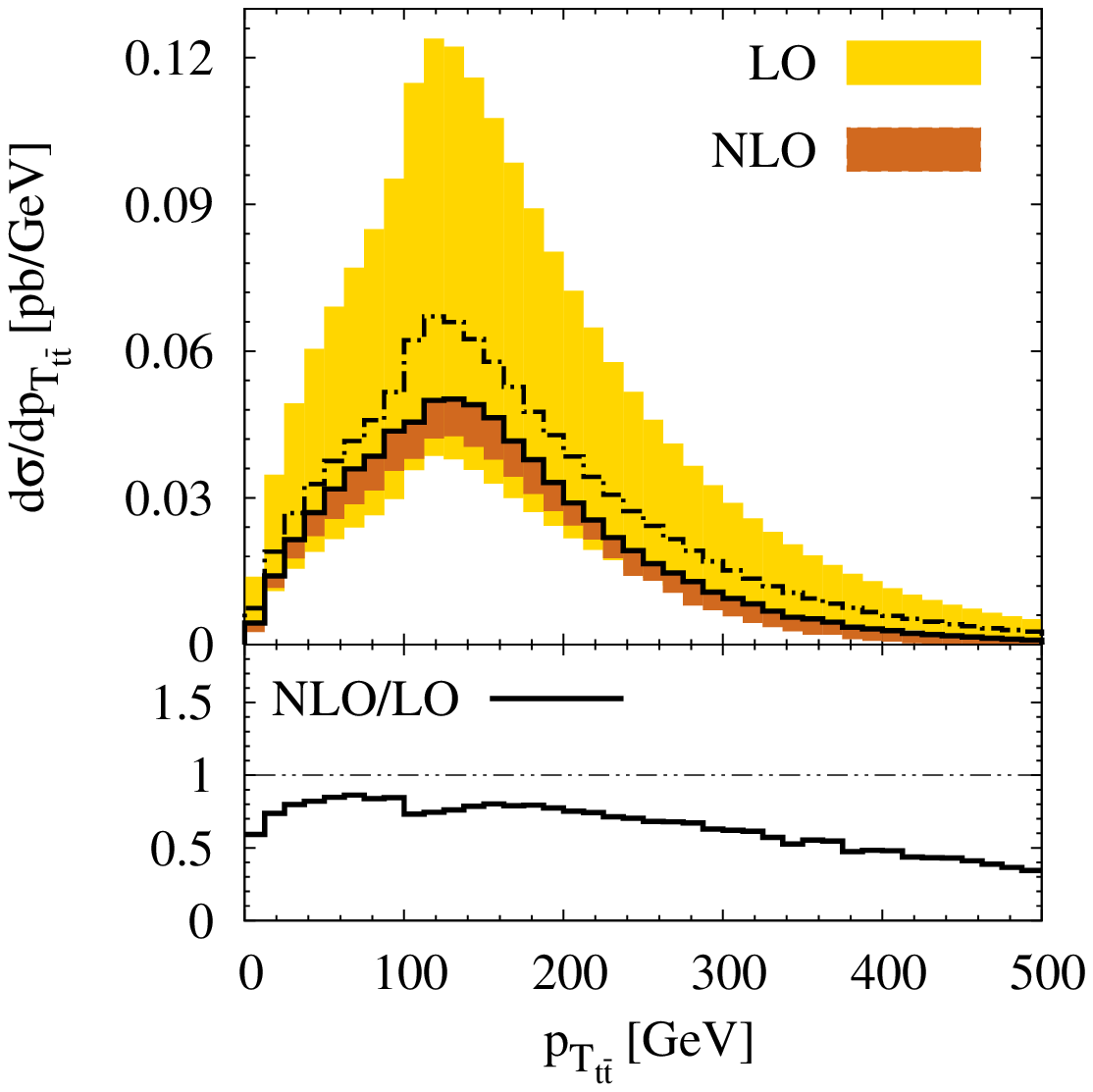}
\caption{\it  \label{fig:lhc3} Differential cross section
  distributions as a function of rapidity (left panel) and  transverse
  momentum (right panel)  of the  $t\bar{t}$ pair  for  $ pp \to t
  \bar{t} jj + X$ production  at the LHC with  $\sqrt{s}= 7
  ~\textnormal{TeV}$. The dash-dotted curve corresponds to the LO,
  whereas the solid one to the NLO result.  The uncertainty bands
  depict scale variation. The lower panels display the differential
  $\cal K$ factor.}
\end{figure*}
\begin{figure*}
\includegraphics[width=0.48\textwidth]{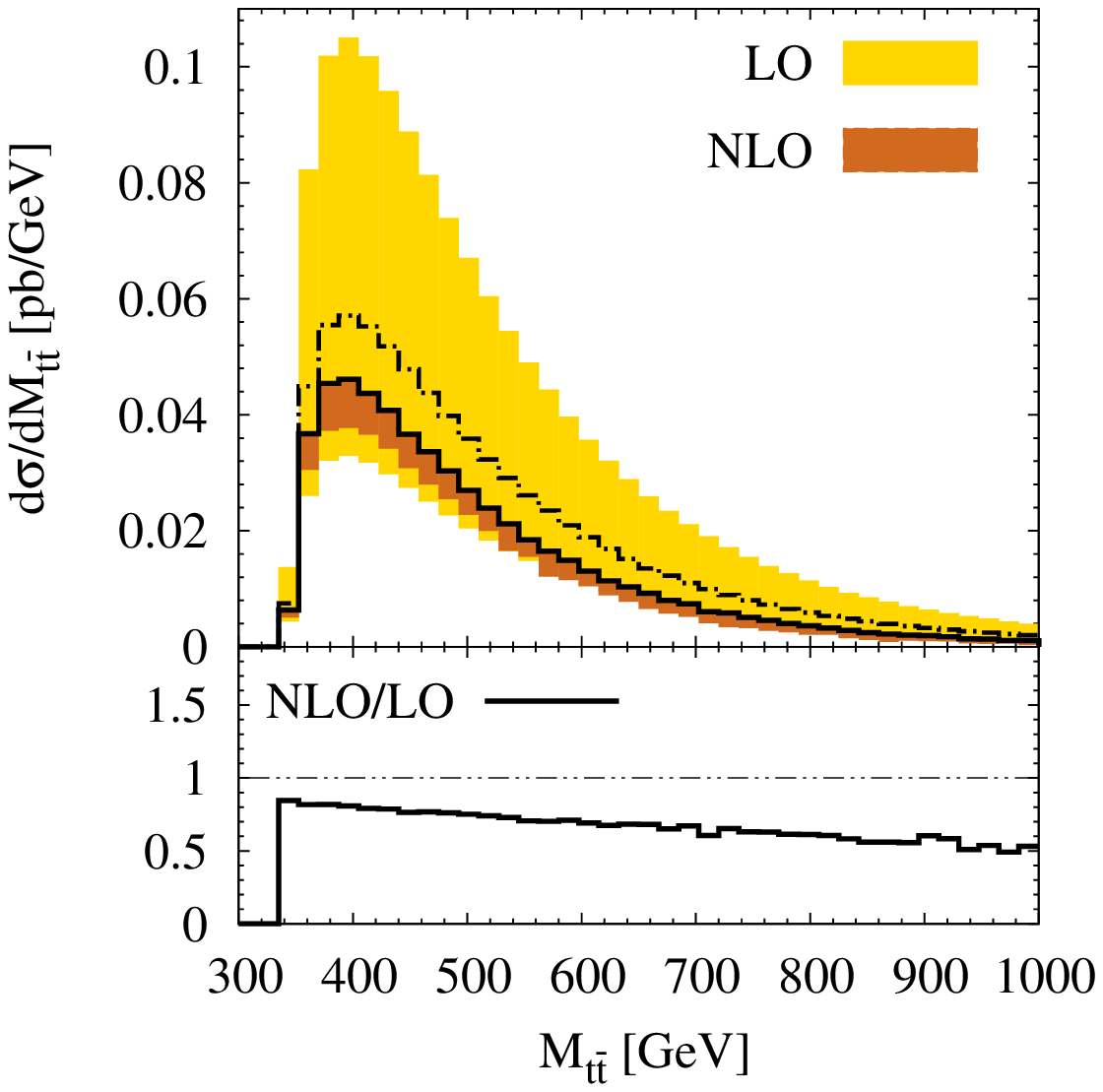}
\includegraphics[width=0.48\textwidth]{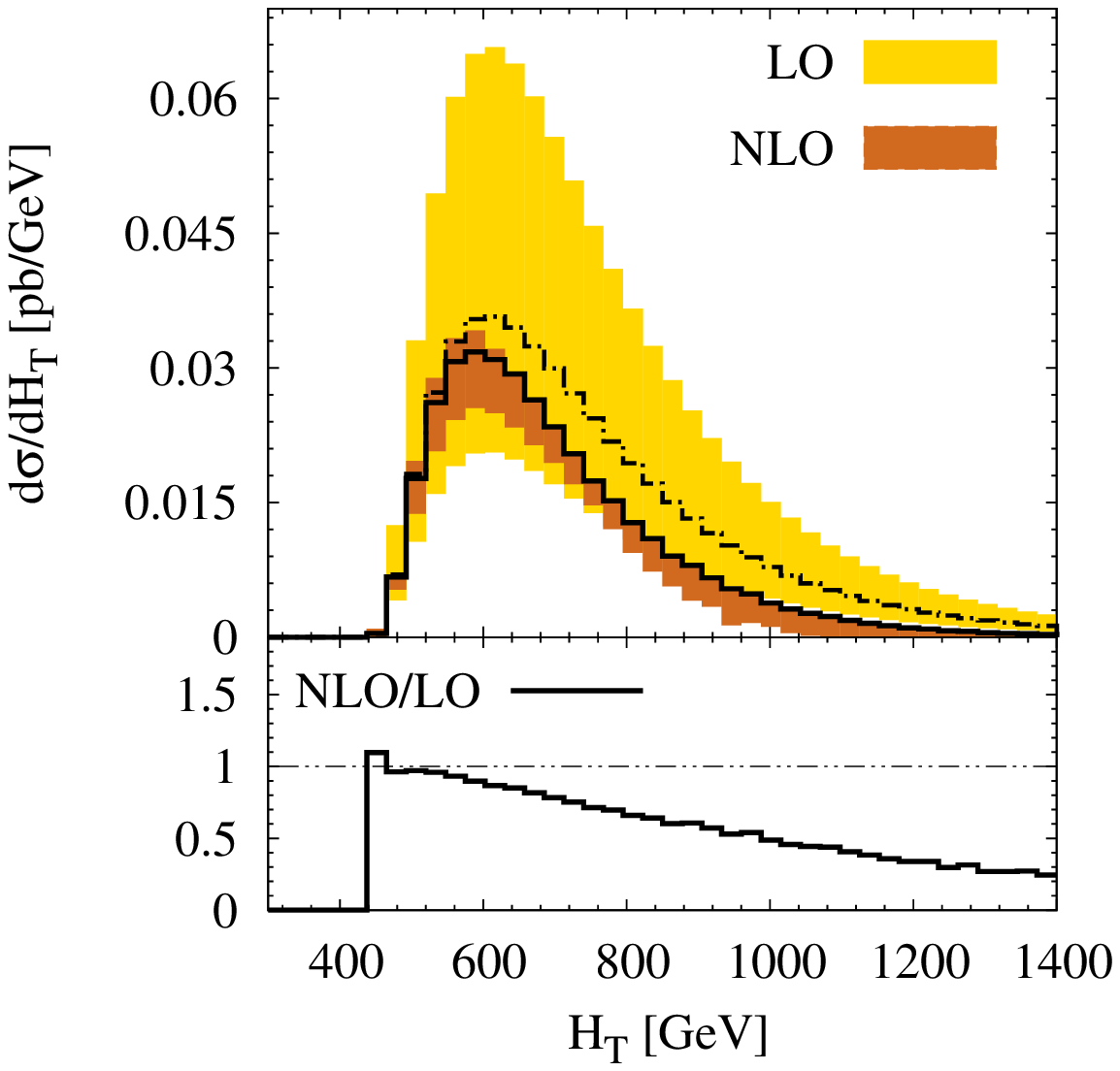}
\caption{\it  \label{fig:lhc4} Differential cross section
  distributions as a function of the invariant mass of the  $t\bar{t}$
  pair (left panel) and of total transverse energy (right  panel)
  for  $ pp \to t \bar{t} jj + X$ production  at the LHC with
  $\sqrt{s}= 7 ~\textnormal{TeV}$. The dash-dotted curve corresponds
  to the LO, whereas the solid one to the NLO result.  The uncertainty
  bands depict scale variation. The lower panels display the
  differential $\cal K$ factor.}
\end{figure*}
\begin{figure*}
\includegraphics[width=0.48\textwidth]{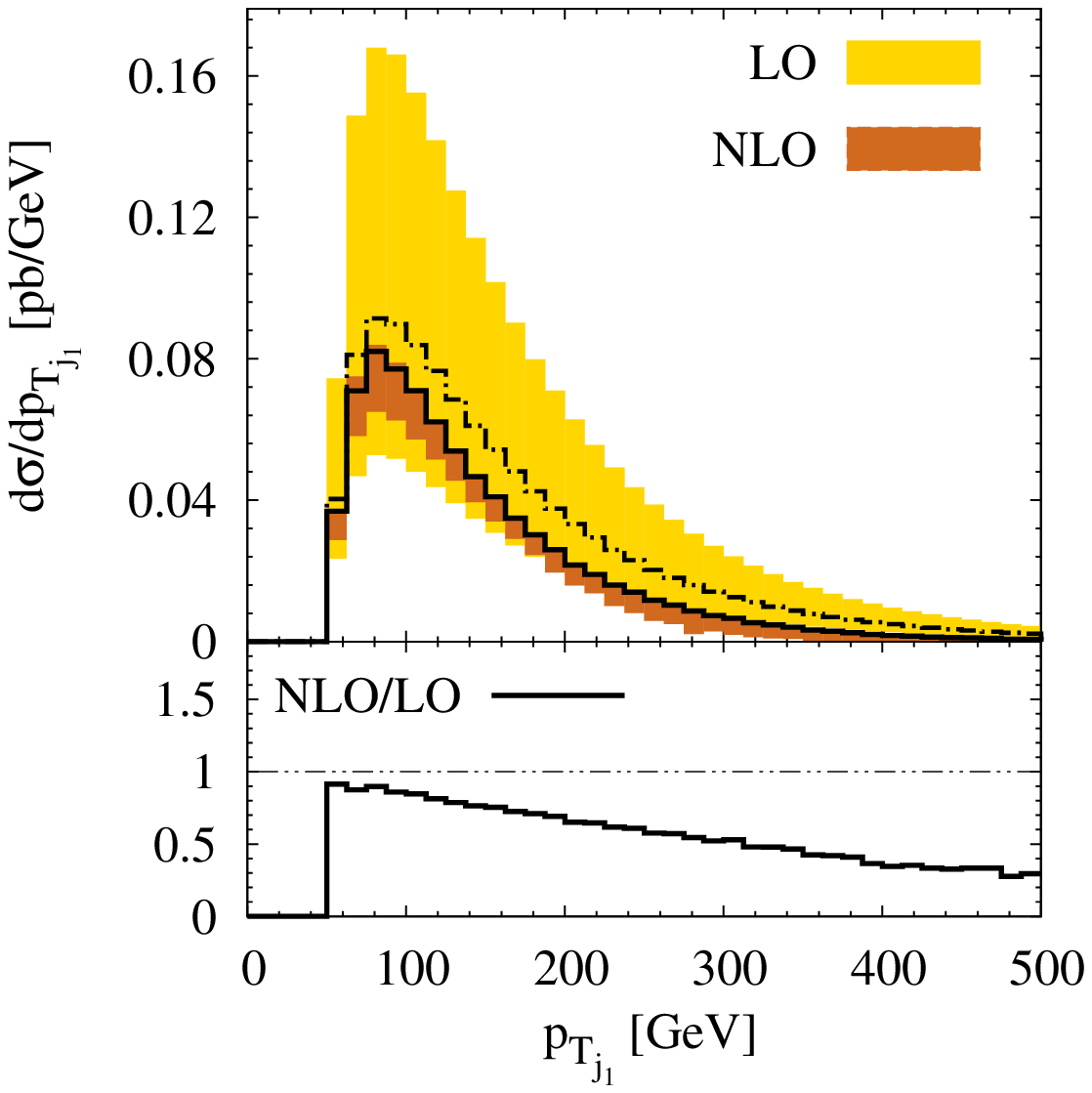}
\includegraphics[width=0.48\textwidth]{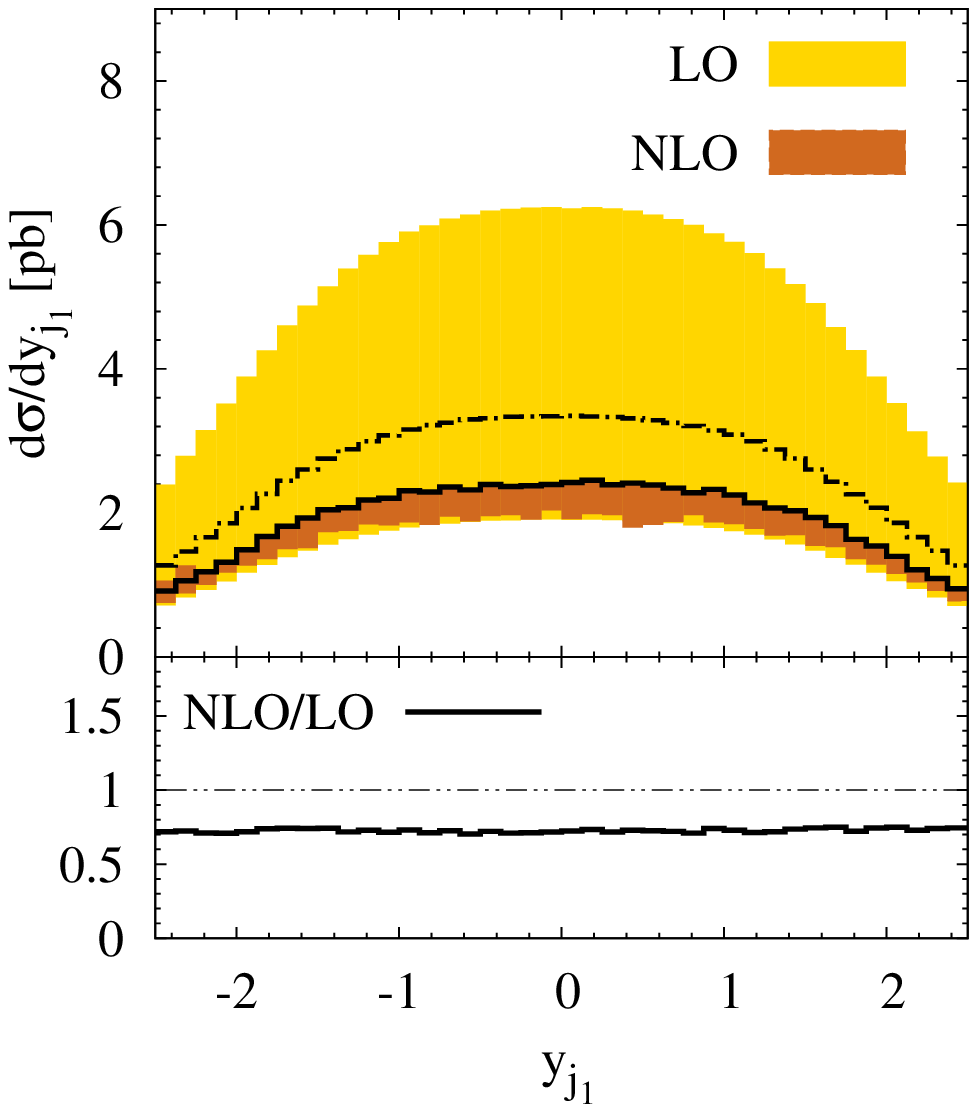}
\caption{\it  \label{fig:lhc5} Differential cross section
  distributions as a function of transverse momentum (left  panel) and
  rapidity (right  panel) of the 1st hardest jet  for  $ pp \to t
  \bar{t} jj + X$ production  at the LHC with  $\sqrt{s}= 7
  ~\textnormal{TeV}$. The dash-dotted curve corresponds to the LO,
  whereas the solid one to the NLO result. The uncertainty  bands
  depict scale variation. The lower panels display the differential
  $\cal K$ factor.}
\end{figure*}
\begin{figure*}
\includegraphics[width=0.48\textwidth]{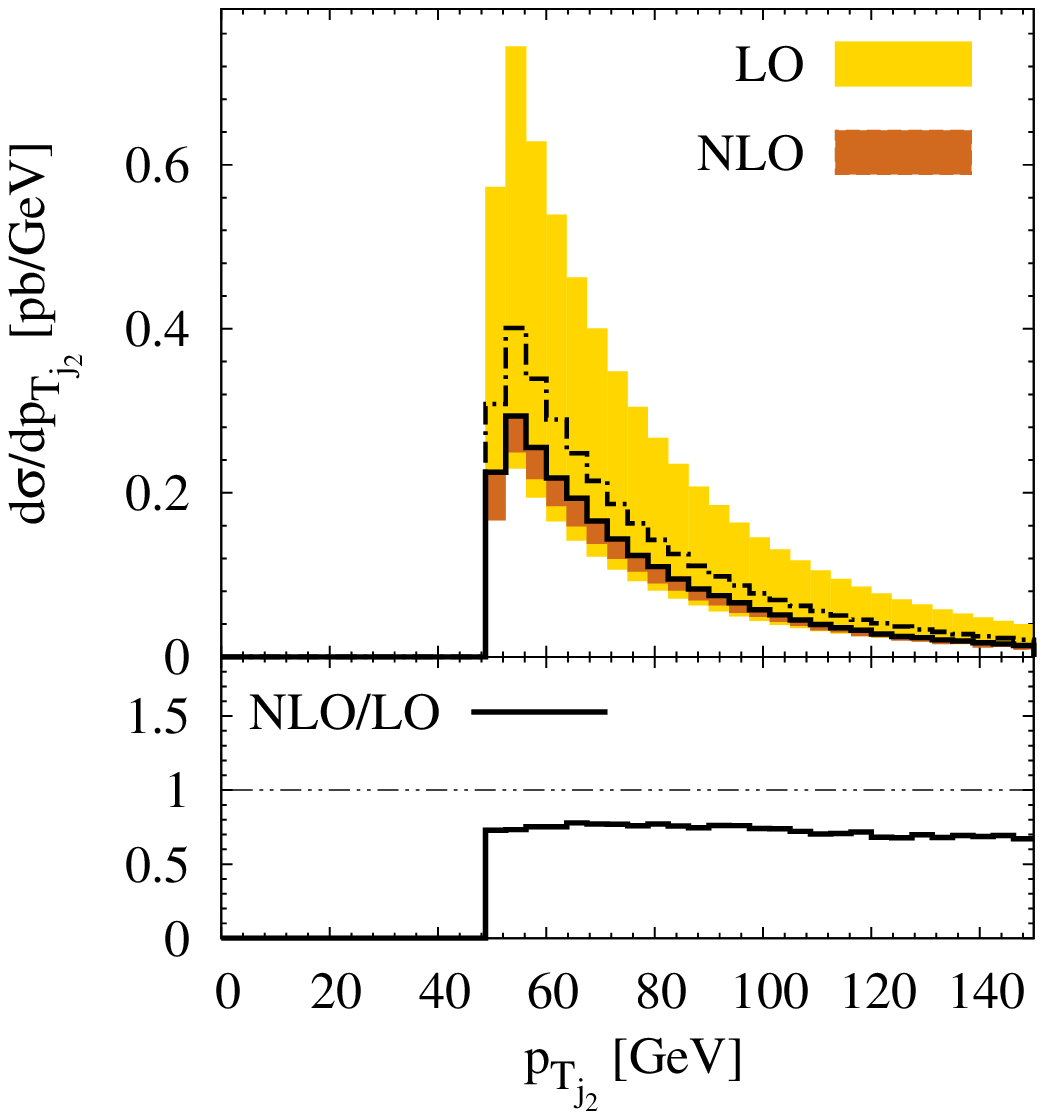}
\includegraphics[width=0.48\textwidth]{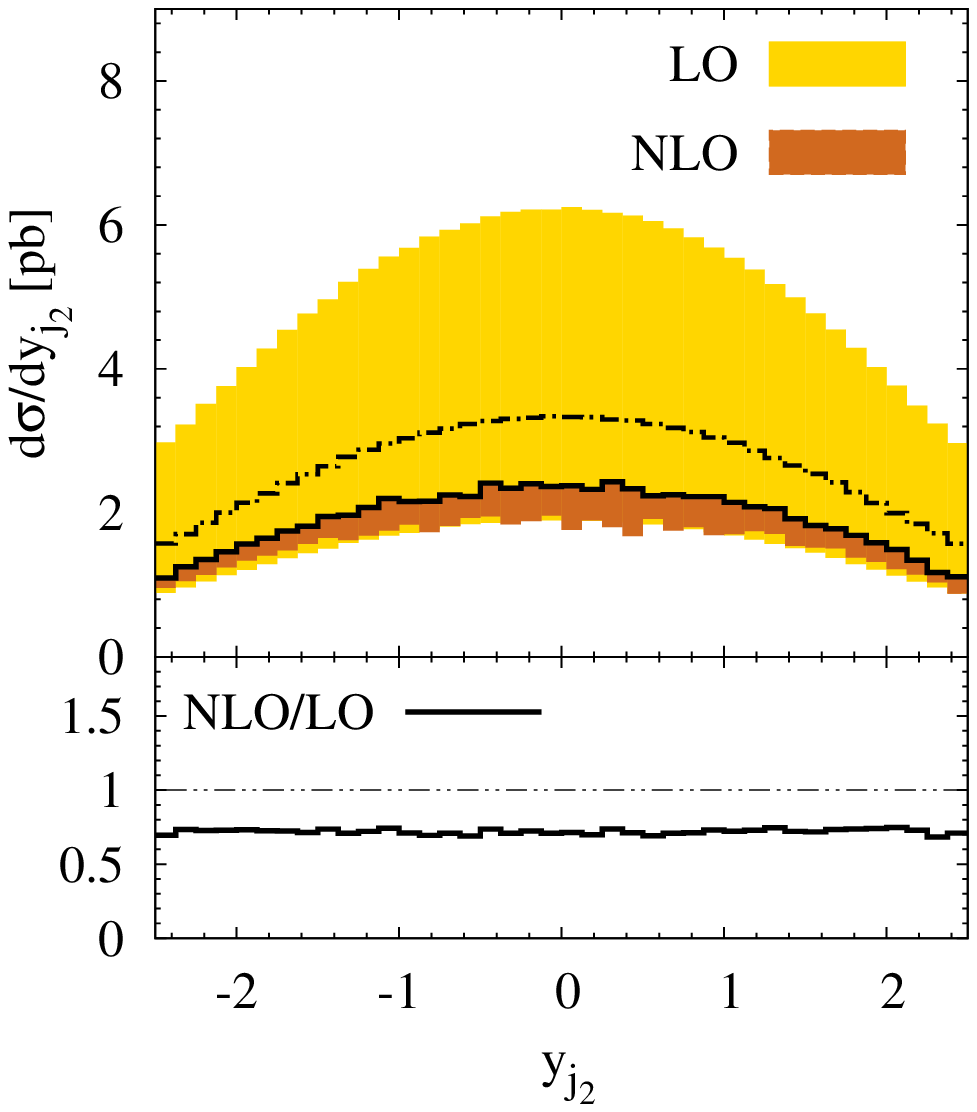}
\caption{\it  \label{fig:lhc6} Differential cross section
  distributions as a function of  transverse momentum  (left panel)
  and rapidity (right panel) of the 2nd hardest jet  for  $ pp \to t
  \bar{t} jj + X$ production  at the LHC with  $\sqrt{s}= 7
  ~\textnormal{TeV}$. The dash-dotted curve corresponds to the LO,
  whereas the solid one to the NLO result. The uncertainty  bands
  depict scale variation. The lower panels display the differential
  $\cal K$ factor.}
\end{figure*}
\begin{figure*}
\includegraphics[width=0.48\textwidth]{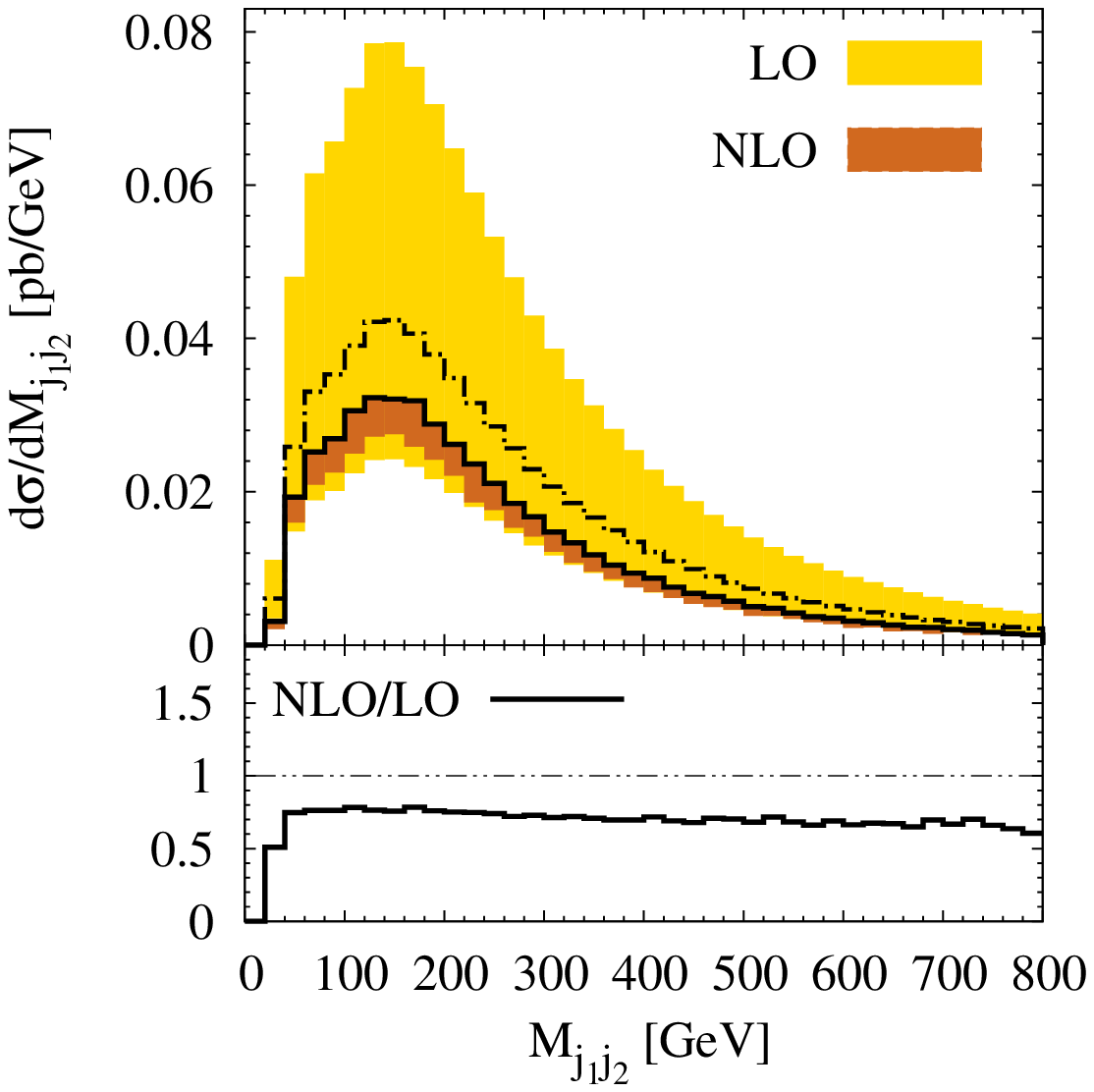}
\includegraphics[width=0.48\textwidth]{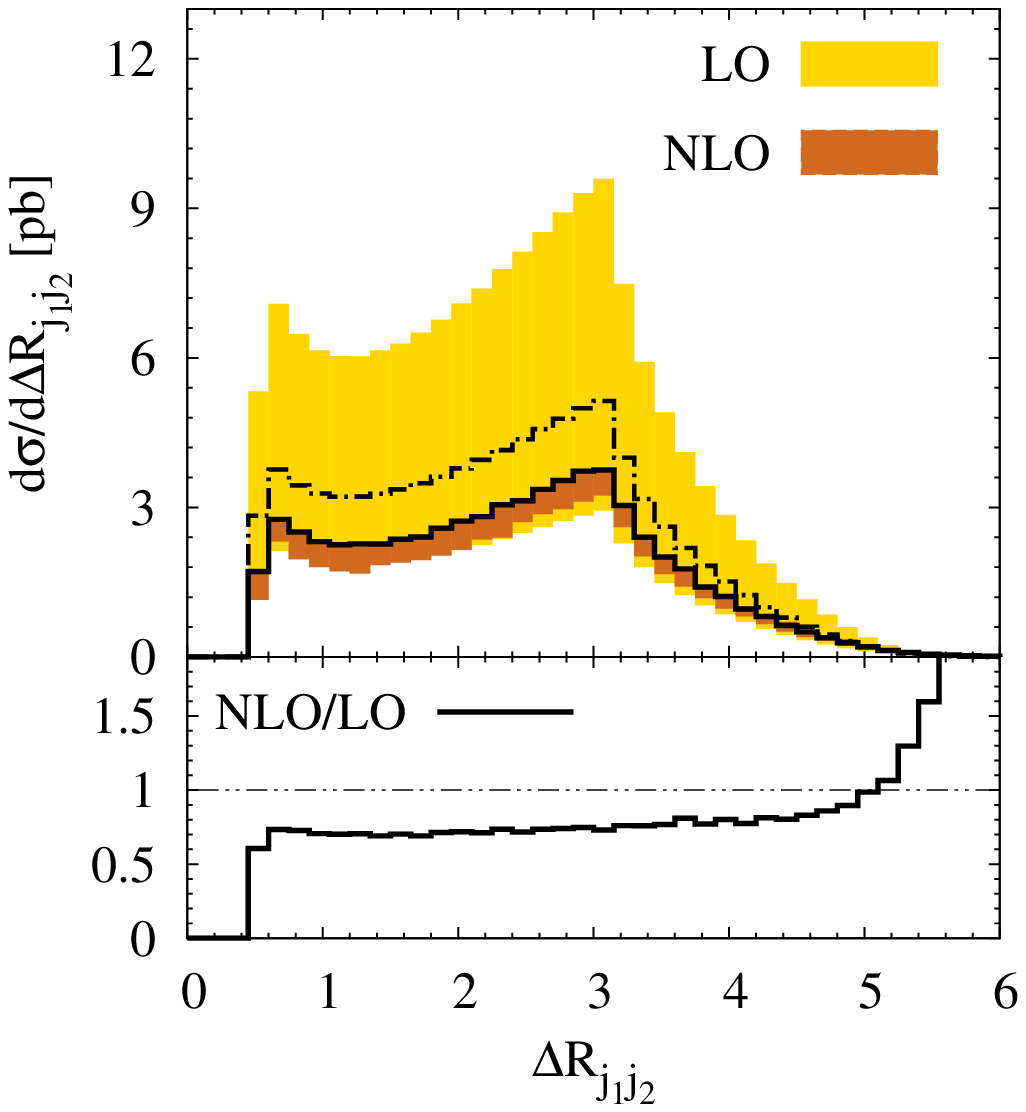}
\caption{\it  \label{fig:lhc7}
 Differential cross section distributions as a function of the
  invariant mass, $m_{jj}$, of the 1st and the 2nd hardest jet and $\Delta
  R_{jj}$ separation  for  $ pp \to t \bar{t} jj + X$ production
  at the LHC with  $\sqrt{s}= 7  ~\textnormal{TeV}$. The dash-dotted
  curve corresponds to the LO, whereas the solid one to the
  NLO result. The uncertainty bands depict scale variation. 
  The lower panels display the differential $\cal K$  factor.}
\end{figure*}
%
We turn now our attention to the LHC results. Let us stress here, that
the cross sections and differential distributions  presented  in this
section  do not coincide with the ones shown in
\cite{Bevilacqua:2010ve}. In the latter publication, a
different center of mass energy,  $\sqrt{s} = 14$ TeV, 
has been assumed.

First, integrated cross sections with the LHC default selection as defined 
in the previous section, are provided. We obtain 
\begin{equation}
\sigma_{\rm LO}^{\rm LHC}(pp\rightarrow t\bar{t}jj)
=13.398^{\, +11.713(87\%)}_{\, -5.788(43\%)} 
~{\rm pb} \,,
\end{equation} 
\begin{equation}
\sigma_{\rm NLO}^{\rm LHC}(pp\rightarrow t\bar{t}jj)
=9.82^{\, -1.48(15\%)}_{\, -1.47(15\%)} 
~{\rm pb} \,.
\end{equation}   
where the central value refers to $\mu_R=\mu_F=m_t$, the upper value
to $\mu=m_t/2$ and the lower value to $\mu=2m_t$.  At the central
value of the scale the total cross section receives  negative and
moderate corrections of the order of $27\%$. Similar to the TeVatron
case, the improvement in the scale  stability at the NLO is
prominent. The NLO cross section is very stable  against changes in the
renormalization and factorization scales, in contrary to the LO
result where an $87\%$ scale dependence is obtained.  

As in the TeVatron  case we have checked the effect coming from the jet
algorithm variation. Results are presented   in Table
\ref{tab:lhc1}. For the jet resolution parameter $R=0.5$ the
difference between $k_T$, $anti-k_T$ and the inclusive Cambridge/Aachen
(C/A) jet finders is within $0.4\%$. With an increase of  $R$ values
to $R=1.0$ the difference moves  upwards  to $1.1\%$ as can be seen
from Table \ref{tab:lhc2}. Moreover, for a higher jet separation cut
$\Delta R_{jj} >1.0$ and a higher jet  resolution parameter $R=1$, NLO
QCD corrections are  reduced down to $-14\%$. Again, the LO result scales
down, this time by $14\%$,  while in the NLO case we have an increase
of the order of $2\%$.  At  LO one parton is equivalent to one jet,  and  the
jet algorithm  with size parameter $R$ is equivalent to the requirement
that any two partons  should be separated by at least $\Delta R_{jj}$
distance. A larger value of  $\Delta R_{jj}$ means smaller cross
sections. At NLO, there can be two  partons in a jet, and jets may
have some structure.  A larger $R$ means that the probability 
of radiation of a parton out of the  area with interjet distance $R$  is
smaller.  This can be translated into  a reduced number of jets with
$p_T$ below the cut threshold and higher  total NLO cross
section. Contrary to the TeVatron case, this  has not been  overcome
here by
the simultaneous increase of the $\Delta R_{jj}$ separation cut  resulting
in higher total NLO cross sections.  At
the  LHC, the two hard jets are more likely to be produced
back-to-back. Therefore, events are mostly concentrated  around
$\Delta R_{jj}=\pi$, which is mildly affected by a change  in $\Delta
R_{jj}$ cut from $0.5$ to $1$. 

Similarly to the TeVatron case, NLO cross sections for two different
values  of the $\alpha_{\rm max}$ parameter are given in Table
\ref{tab:lhc3}. The  independence of the final result on $\alpha_{\rm
  max}$ is  obtained again  at the permil level. 

In Table \ref{tab:lhc4} predictions for LO and NLO cross sections are
presented with the default LHC selection, however, the transverse
momentum  cut of the two hard jets is varied between $50$ GeV and $125$
GeV. Within a $50-100$ GeV range corrections are quite stable,
changing the ${\cal K}$ factor  by less than $9\%$. For $125$  GeV
cut, we observed a somehow higher rise of NLO QCD corrections up to
$-40\%$.  Compared to the total NLO $t\bar{t}$ cross section, 
\begin{equation}
\sigma_{\rm NLO}^{\rm LHC}(pp\rightarrow t\bar{t}) =158.1^{\,
  +19.5(12\%)}_{\, -21.2(13\%)}  ~{\rm pb} \,,
\end{equation} 
we discover that for the
smallest  $p_{T_j}$ cut of $50$ GeV the $t\bar{t}jj$ events represent
only $6\%$ of the total cross section. The fraction is decreased to
$3\%$, $1\%$ and $0.6\%$ for $75$ GeV, $100$ GeV and $125$ GeV
respectively.
%
\subsubsection{Differential cross sections}
%
In the following,  we turn our attention to the size of the NLO QCD
corrections to the differential distributions at the LHC with
$\sqrt{s}=7 ~{\rm TeV}$.  For the LHC, a $pp$ collider, the shapes of
some distributions are quite different as compared to the TeVatron
case.  Top quark  production at the LHC, for example,  is
forward-backward symmetric in the laboratory frame as a consequence of
the symmetric initial state. Thus, ${\cal A}_{\rm FB}^{t}={\cal
  A}_{\rm FB}^{\bar t}=0$. This can be observed in  Figure
\ref{fig:lhc1}, where top and anti-top quark
rapidity distributions, symmetric around $y_t$, are plotted at LO and NLO.  Distributions get
broadened, moving from the TeVatron to the LHC case, as can be
expected from an increased scattering  energy.  On the other hand,
the distribution in the averaged transverse momentum, $p_{T_t}$, of the top
and anti-top, shown in Figure \ref{fig:lhc2},  becomes harder
during this transition. Again, the dash-dotted  curve corresponds to
the LO,  whereas the solid one to   the NLO result.  At LO the bands
correspond to the choice of the renormalization and factorization
scale $\mu=m_t/2$ and $\mu=2m_t$.  In case of NLO they are determined
out of the following set $\mu=\left[m_t/2,m_t,2m_t\right]$.  More
precisely for each histogram, bands are constructed bin-by-bin, by calculating
maximal and minimal values using the above three scales. Since the
scale dependence of the NLO cross section at the LHC is a 
function that is not monotonic and the true maximum is somewhere close
to the central value of the scale, $m_t$, this seems to be the only
meaningful approach, if one is interested in plotting scale uncertainty
bands without making a very costly scan.  Let us emphasize here, that
the scale variation is, by no means a 
rigorous way to  estimate the theoretical uncertainty. At best, it
might only give an  indication of the full uncertainty  which is due
to the not yet calculated  higher order corrections.  The lower
panels display the  differential  $\cal K$ factor.  Angular
distributions exhibit negative corrections of the order of $30\%$ in
the central rapidity range and they are decreased to $20\%$ at the
forward and backward parts of the spectrum. As for the averaged
transverse momentum  distributions of the top and anti-top up to
$60\%$ distortions are observed.

In Figure \ref{fig:lhc3} and Figure \ref{fig:lhc4}, differential cross
section distributions as function of  rapidity, $y_{t\bar{t}}$,
transverse momentum, $p_{T_{t\bar{t}}}$, and  the invariant mass,
$m_{t\bar{t}}$, of the $t\bar{t}$ pair are given. Furthermore, the
total transverse energy of the  system, $H_T$, is shown there.
The rapidity distribution  receives negative, but rather  stable $25\%$
corrections. Unlike      $p_{T_{t\bar{t}}}$, $m_{t\bar{t}}$ and $H_T$
spectrums, where distortions of the order of  $50\%$, $35\%$ and
$80\%$ respectively, are visible.

Subsequently, in Figure \ref{fig:lhc5} and in  Figure \ref{fig:lhc6}
differential cross section distributions as a function of  transverse
momentum, $p_{T_{j}}$, and rapidity, $y_{j}$, of the 1st and the 2nd
hardest jets are plotted.  As anticipated for rapidity distributions, in
both cases negative and quite  stable corrections of the order of
$30\%$ appear. On the other hand, for the $p_{T_{j_1}}$
spectrum, $60\%$ distortions are obtained as compared  to only $10\%$
in  $p_{T_{j_2}}$ case.

Eventually, differential cross section distributions as a function of
the dijet invariant mass, $m_{jj}$, are given in Figure \ref{fig:lhc7}.
Also shown in Figure \ref{fig:lhc7} is the  separation in $\Delta
R_{jj}$ between two jets  ordered in hardness.  In both cases, rather
constant $-30\%$ corrections can be reached. Again, fluctuations
visible at the tails of the $\Delta R_{jj}$ distribution are caused by
limited  statistics of the Monte Carlo integration. As we can see  at
the LHC, the two hard jets  are more likely to be produced
back-to-back, leading to a more  peaked distribution around $\pi$.

Generally,  we can say that for a fixed scale $\mu=\mu_R=\mu_F=m_t$ at the
LHC, the NLO QCD corrections are always negative and protean. In case of
angular distributions, $y$, $\Delta R$, and the invariant mass distribution of
the $t\bar{t}$ pair,  rather steady $20\%-30\%$ corrections are visable. In
the transverse momentum  distributions, on the other hand, they introduce up to
$60\%$ distortions. In addition, in the case of the total transverse
energy, $H_T$, even $80\%$ deformations can be  observed. 
%
\section{Conclusions}
\label{sec:3}
In this paper, we have presented a computation of the NLO QCD
corrections to $t\bar{t}$ pair production in association with two hard jets
at the TeVatron and  the LHC.  The total cross sections and their scale
dependence, as well as several differential distributions have been
given.  The NLO QCD corrections 
reduce the scale uncertainty of the total cross sections and of the
differential distributions compared to LO calculations, which can
only provide qualitative predictions.  

In case of the
TeVatron, the forward-backward asymmetry of the top quarks has been
calculated for the first time. We have found that, with inclusive
selection cuts, the forward-backward asymmetry  amounts to  ${\cal
  A}_{\rm FB, LO}^{\rm t} = -10.3\% $ at LO and  ${\cal A}_{\rm FB,
  NLO}^{\rm t}  = -4.6 \%$ at NLO. 

Furthermore, the impact of the NLO
QCD corrections on integrated cross sections at the TeVatron is
moderate, of the order of $-24\%$. A the LHC we have obtained NLO QCD
corrections at the level of  $-27\%$. In both cases, NLO QCD corrections
are reduced substantially, if a higher jet  resolution parameter $R$ if
chosen instead. Changing $R$ from $R=0.4$ to $R=0.8$ in the TeVatron
case  and from $R=0.5$ to $R=1.0$ in the LHC case decreases  NLO QCD
corrections down to  $-16\%$ and $-14\%$ respectively.  A study of the
scale dependence of our NLO predictions indicates that the residual
theoretical uncertainty due to higher order corrections is $21\%$ for
the TeVatron and $15\%$ for the LHC.  

As a further matter, at the
TeVatron employing a fixed scale $\mu=\mu_R=\mu_F=m_t$, the NLO
corrections to differential distributions are negative and rather substantial.
They do not simply rescale the LO shapes, but induce distortions up to
$60\%$. Also in the LHC case the NLO QCD corrections to the
distributions  are always negative and protean. In the worst case even
$80\%$ deformations can be  observed.  

Finally, our paper demonstrates the utility of the off-shell methods
and the OPP reduction  procedure, as implemented in the \textsc{Helac-NLO}
system, in computing NLO QCD  corrections for processes of
phenomenological interest at hadron colliders.  The system consists
of, \textsc{CutTools} and \textsc{Helac-1Loop} which handle the
virtual corrections  and \textsc{Helac-Dipoles} for the real emission
contributions. For the  phase space integration the \textsc{Kaleu}
package is used and  results  are cross checked with the help of the
\textsc{Phegas} phase space generator.  Moreover, \textsc{OneLOop}  is
deployed  for the evaluation of the one-loop  scalar functions.  With this
framework we can calculate NLO QCD corrections to the remaining
multiparticle background and signal processes which are relevant  in
the ongoing analyzes  at the TeVatron and at the LHC. Besides, we
are prepared to compare  NLO QCD predictions with already
available data.

%
\section{Appendix}
\label{sec:4}
In order to facilitate a comparison to our calculation, we provide
values of the one-loop virtual corrections to the squared matrix
elements for one phase space point which is given in Table
\ref{tab:appendix1}. Virtual corrections come from the interference of
the tree level and one-loop amplitudes, summed over all colors and
spins, and for $n_f=5$ massless quark flavors. As mentioned before we
describe the running of the strong coupling constant with two-loop
accuracy, including five active flavors which corresponds to
$\alpha_s(m_t) = 0.109403691593462$ where  $m_t = 173.3$ GeV. In Table
\ref{tab:appendix2}  we present numbers for all eight
subprocesses. The quantities presented in Table \ref{tab:appendix2}
correspond to the renormalized virtual one-loop correction in  the 't
Hooft-Veltman \cite{'tHooft:1972fi} version of the dimensional
regularization scheme.  The remaining singularities in the dimensional
regularization parameter $d=4-2\epsilon$ arise from the virtual soft
and collinear singularities in the one-loop amplitudes. Additionally
coefficients for color and spin summed results  for the  ${\cal
  I}$-operator of the dipole subtraction function as defined in
\cite{Catani:2002hc}  are presented. We can observe numerical
cancellations of infrared divergences between the ${\cal I}$-operator
and the corresponding virtual contributions at least up to 10 digits.
%
\begin{table*}
  \caption{\it \label{tab:appendix1} The set of momenta for  $ 1\,2\, \to
    t\bar{t} \,3\,4$ process with the notation $(p_x,p_y,p_z,E)$ and all the
    components given in GeV. The top quark mass is set to $m_t=173.3$ GeV.}
\begin{ruledtabular}
  \begin{tabular}{ccccc}
   \textsc{Parton}  &\textsc{$p_x$}  &
      \textsc{$p_y$}    &
     \textsc{$p_z$} & \textsc{E} \\
&&&&\\
\hline
$p_1$ & 0 & 0 & 2424.746502697520 & 2424.746502697520\\
$p_2 $& 0 & 0 & -2424.746502697520 & 2424.746502697520\\
$p_t $&-715.3340594013138 & -475.1625187999429 &101.1925816377932 &
881.9042263139404  \\
$p_{\bar{t}} $ &-24.14783219601748 & -6.328336607570634 & 295.5085181344487
& 343.4841188963524  \\
$p_3 $ & 21.87826794850173 & 1000.411563795763 &  1341.334408905234 & 
1673.463460042633  \\
$p_4 $ & 717.6036236488295 & -518.9207083882493 & -1738.035508677476 & 
1950.641200142114 \\
  \end{tabular}
\end{ruledtabular}
\end{table*}
%
\begin{table*}
  \caption{\it \label{tab:appendix2}  Numerical values of the virtual
    corrections to the squared matrix elements at the phase space point given
    in Table \ref{tab:appendix1}  and
    $\mu_R=\mu_F=m_t$ for all the partonic subprocesses for the
    $t\bar{t}jj$ production at a hadron collider. We give the finite parts
    along with the coefficients of the poles in $\epsilon$.  Additionally
    coefficients for color and spin summed results  for the  ${\cal
      I}$-operator are presented. }
\begin{ruledtabular}
  \begin{tabular}{cccc}
     \textsc{Subprocess}  &
     \textsc{$1/\epsilon^2$}    &
     \textsc{$1/\epsilon$} & \textsc{$1/\epsilon^0$} \\
&&&\\
\hline
 $gg\to t\bar{t}gg $& -4.759403785663033$\times 10^{-9}$ 
&\,\,1.868033206671699$\times 10^{-8}$
 &-3.727507250164706$\times 10^{-8}$\\
 ${\cal I}(\epsilon)$& \,\,4.759403785663084$\times 10^{-9}$ 
& -1.868033206665652$\times 10^{-8}$ &\\
&&&\\
 $gg\to t\bar{t}u\bar{u} $& -1.243597961079720$\times 10^{-11}$ &
\,\,4.205081694102065$\times 10^{-11}$& -8.435931194456964$\times 10^{-11}$\\
 ${\cal I}(\epsilon)$&\,\,1.243597961079740$\times 10^{-11}$& 
-4.205081694097688$\times 10^{-11}$&\\
&&&\\
 $u\bar{u}\to t\bar{t}gg $&-3.408532130149515$\times 10^{-11}$ & 
\,\,1.296674677886268$\times 10^{-10}$ &-2.908105638419785$\times 10^{-10}$ \\
 ${\cal I}(\epsilon)$&\,\,3.408532130149502$\times 10^{-11}$ 
&-1.296674677886843$\times 10^{-10}$&\\
&&&\\
 $u\bar{u}\to t\bar{t}u\bar{u} $& -5.610006218609934$\times 10^{-10}$ & 
\,\,3.140716653723494$\times 10^{-9}$ & -6.264131939313179$\times 10^{-9}$\\
 ${\cal I}(\epsilon)$ &\,\,5.610006218609923$\times 10^{-10}$
 &  -3.140716653723457$\times 10^{-9}$&\\
&&&\\
 $ug\to t\bar{t}ug $&-2.178331428688374$\times 10^{-9}$
& \,\,9.512663847206611$\times 10^{-9}$&
-1.925455596882764$\times 10^{-8}$\\
 ${\cal I}(\epsilon)$&\,\,2.178331428688390$\times 10^{-9}$ &
-9.512663847144086$\times 10^{-9}$&\\
&&&\\
 $gu\to t\bar{t}ug$&-5.212043686490916$\times 10^{-11}$ 
&\,\,1.749451737240981$\times 10^{-10}$ 
&-3.143378807927551$\times 10^{-10}$ \\
 ${\cal I}(\epsilon)$&\,\,5.212043686491147$\times 10^{-11}$ 
&-1.749451737238111$\times 10^{-10}$&\\
&&&\\
$ud\to t\bar{t}ud$ &-5.488979123409352$\times 10^{-10}$
&\,\,3.020380070883476$\times 10^{-9}$
&-5.979795310554041$\times 10^{-9}$
\\
 ${\cal I}(\epsilon)$& \,\,5.488979123409385$\times 10^{-10}$& 
-3.020380070883462$\times 10^{-9}$&\\
&&&\\
$u\bar{u}\to t\bar{t}d\bar{d}$ &-2.663317234536143$\times 10^{-13}$
&\,\,─1.278242588447108$\times 10^{-12}$
&-4.283899547423337$\times 10^{-12}$
\\
 ${\cal I}(\epsilon)$&\,\,2.663317234536076$\times 10^{-13}$  
&-1.278242588446499$\times 10^{-12}$&\\
  \end{tabular}
\end{ruledtabular}
\end{table*}
%
\begin{acknowledgments}
The calculations have been performed on the
Grid Cluster of the Bergische Universit\"at Wuppertal, financed
by the Helmholtz - Alliance ``Physics at the Terascale'' and the BMBF.

M. Czakon was supported by the Heisenberg and by
the Gottfried Wilhelm Leibniz Programmes of the Deutsche
Forschungsgemeinschaft and M. Worek by the Initiative and Networking Fund of
the Helmholtz Association, contract HA-101 (``Physics at the Terascale'').

M. Worek would also like to thank German Rodrigo for helpful discussions.

\end{acknowledgments}

\end{document}